\documentclass{article}
\usepackage[numbers]{natbib}
\usepackage[utf8]{inputenc}
\usepackage[english]{babel}
\usepackage{amssymb}

\usepackage[letterpaper,top=2cm,bottom=2cm,left=3cm,right=3cm,marginparwidth=1.75cm]{geometry}

\usepackage[table,xcdraw,dvipsnames]{xcolor}
\usepackage{xspace}
\usepackage{qcircuit}
\usepackage{amsmath,braket}
\usepackage{graphicx} % Required for \resizebox
\usepackage{booktabs}
\usepackage{pdflscape}
\usepackage{tcolorbox}

\newcommand{\perfromacerateincrease}{0.12\%\xspace} % Results 20/06/2024
\usepackage{booktabs}

\newcommand{\code}[1]{\texttt{#1}}
\usepackage{graphicx}
\usepackage{subcaption}
\usepackage{authblk}
\usepackage{tikz}
\usepackage{multirow}

\newcommand{\probP}{\text{I\kern-0.15em P}}

\usepackage[colorlinks=true, allcolors=blue]{hyperref}

\usepackage{orcidlink}
\newcommand{\sectref}[1]{%
  \hyperref[#1]{Section~\ref{#1}}%
}
\newcommand{\subsectref}[1]{%
  \hyperref[#1]{Subsection~\ref{#1}}%
}

\newcommand{\tool}{\texttt{AccelerQ}\xspace} % tool name
\newcommand{\adapt}{ADAPT-QSCI\xspace}
\newcommand{\qcels}{QCELS\xspace}
\newcommand{\filesTrainAdapt}{33\xspace} % 9/8/2024
\newcommand{\adaptrecA}{60\xspace} % 4-qubit
\newcommand{\adaptrecB}{750\xspace} % 6-qubit
\newcommand{\adaptrecC}{500\xspace} % 7-qubit
\newcommand{\adaptrecD}{500\xspace} % 8-qubit
\newcommand{\adaptrecE}{1500\xspace} % 10-qubit
\newcommand{\adaptrecF}{450\xspace} % 12-qubit
\newcommand{\adaptrecG}{800\xspace} % 14-qubit
\newcommand{\adaptrecH}{200\xspace} % 16-qubit
\newcommand{\QCELSrecA}{60\xspace} % 4-qubit
\newcommand{\QCELSrecB}{750\xspace} % 6-qubit
\newcommand{\QCELSrecC}{500\xspace} % 7-qubit
\newcommand{\QCELSrecD}{500\xspace}  % 8-qubit
\newcommand{\QCELSrecE}{1000\xspace} % 10-qubit
\newcommand{\QCELSrecF}{6600\xspace} % 12-qubit
\newcommand{\QCELSrecG}{2100\xspace} % 14-qubit
\newcommand{\QCELSrecH}{3000\xspace} % 16-qubit
\newcommand{\gpuspec}{a single GPU core, NVIDIA 12GB PCI P100 GPU, 12 GB VRAM, running Ubuntu 22.04.4 LTS\xspace} % Cloud-lab c240g5-110219 node
\newcommand{\specdeploy}{an ARM M2 Mac running Ubuntu 22.04.4 with 8 GB RAM\xspace} % experiments Aug 16, 2024
\newcommand{\qcelstrainsize}{4868 MB\xspace} % TODO
\newcommand{\adapttrainsize}{757 MB\xspace} % 66 items, totalling 757.0 MB, 16 AUG, 2024
\newcommand{\qcelstrainsizemodel}{2.86 MB\xspace} % TODO
\newcommand{\adapttrainsizemodel}{1.1 MB\xspace} % 1.1 MB (1,093,386 bytes) 16 AUG, 2024

\newcommand{\adapttenMAE}{6.28776\xspace} % 6.2877638154543645
\newcommand{\adapttenMSE}{129.512\xspace} % 129.51205875315478 
\newcommand{\adapttwentyMAE}{6.37912\xspace} % 6.37912112280173
\newcommand{\adapttwentyMSE}{122.659\xspace} % 122.65932318752385

\newcommand{\QCELStenMAE}{9.05254\xspace} % 9.052541815752853
\newcommand{\QCELStenMSE}{258.031\xspace} % 258.0303998056656 
\newcommand{\QCELStwentyMAE}{8.25527\xspace} % 8.255266322922873
\newcommand{\QCELStwentyMSE}{145.092\xspace} % 145.09232645760397

\title{Accelerating Quantum Eigensolver Algorithms \\ With Machine Learning}
\author[1]{Avner Bensoussan\orcidlink{0009-0007-3285-9468}}
\author[2]{Elena Chachkarova\orcidlink{0000-0003-2857-5570}}
\author[1]{Karine Even-Mendoza\orcidlink{0000-0002-3099-1189}}
\author[1]{\\Sophie Fortz\orcidlink{0000-0001-9687-8587}}
\author[2]{Connor Lenihan\orcidlink{0000-0003-1885-2941}}

\affil[1]{Informatics, NMES, King's College London, England, UK}
\affil[2]{Physics, NMES, King's College London, England, UK}

\date{}

\begin{document}
\maketitle

%%%%%%%%%%%%%%%%%%%%%%%%%%%%%%%%%%%%%%%%%%%%%%%%%%%%%%%%%%
\begin{abstract}
%In this paper, we describe our experience and efforts to tackle the \textit{Quantum Algorithm Grand Challenge} (QAGC2024)--putting together a team of quantum physics and software engineer scientists--to accelerate the Variational Quantum Eigensolver (VQE) algorithms and quantum eigensolvers in general. 
%% Details about the challenge
%The challenge aimed to accelerate 28-qubit Hamiltonian ground state energy calculation, on NISQ devices, using limited resources (\textit{e.g.,} the number of shots of $10^7$, timeout of $6 \times 10^5$ seconds).

In this paper, we explore accelerating Hamiltonian ground state energy calculation on NISQ devices.
We suggest using search-based methods together with machine learning to accelerate quantum algorithms, exemplified in the Quantum Eigensolver use case. 
We trained two small models on classically mined data from systems with up to 16 qubits, using \code{XGBoost}'s Python regressor.
We evaluated our preliminary approach on 20-, 24- and 28-qubit systems by optimising the Eigensolver's hyperparameters.
These models predict hyperparameter values, leading to
a \perfromacerateincrease{} reduction in error 
when tested on 28-qubit systems. However, due to inconclusive results with 20- and 24-qubit systems, we suggest further examination of the training data based on Hamiltonian characteristics. In future work, we plan to train machine learning models to optimise other aspects or subroutines of quantum algorithm execution beyond its hyperparameters.
\end{abstract}

\section{Introduction}
%\all{Challenges:
%We discuss our approach, its generalisation, and the challenges we encountered while integrating software engineering tools with quantum physics methods.}

Modern-day \textit{Noisy Intermediate-Scale Quantum} (NISQ) devices are the current state-of-the-art in \textit{Quantum Computing} (QC) characterised as noisy with an intermediate scale in terms of the number of qubits they have \cite{preskill_quantum_2018,bharti_noisy_2022,ACAMPORA202316}. 
These devices have not yet evolved to support fault-tolerant calculations with a large enough number of qubits to achieve quantum advantage. However, these already enable quantum researchers and engineers to develop, optimise and test various quantum algorithms. 
In this interim time of intensive hardware development, robust and innovative quantum algorithms \cite{lau_nisq_2022} provide a way to tackle the limitations of the current quantum processors. One of the most promising applications of QC is in the field of quantum chemistry for molecular simulations, structure design, drug design, and more \cite{quantum-computational-chemistry, qc-review, qc-molecules-application}. 
A prominent example of a class of quantum algorithms that is rapidly evolving in the field of quantum chemistry is 
the \textit{Variational Quantum Eigensolver} (VQE) \cite{tilly_variational_2022}, 
which aims at estimating the ground state energy of Hamiltonians. 
Different types of VQE algorithms are being developed and tailored for various systems.

Whilst taking part in the Quantum Algorithm Grand Challenge \cite{QAGC2024} organised by QunaSys\footnote{QunaSys is a technology company aiming to achieve QC's full potential through algorithm development into product engineering} \cite{QunaSys}, we tackle an instance of quantum algorithms for eigensolving, designed to estimate the eigenvalues and eigenvectors of a given Hamiltonian to find its ground state energy. Our approach focuses on applying machine learning algorithms to explore and optimise quantum algorithms to achieve potential advancements in quantum eigensolving and improve the performance and accuracy of quantum algorithms for larger system setups that are currently too complex to tackle.
Our first attempt at the problem was to find a methodology to select the best hyperparameters for the \adapt{} quantum algorithm \cite{QSCI2023} that minimises the runtime and the final result for the ground state energy through using machine learning techniques trained on smaller systems \cite{QAGC_submission}. This method is not specifically tailored to \adapt{} and can be applied to any \textit{Quantum Eigensolver}-based algorithm. 
Next, we applied the \textit{Quantum Complex Exponential Least Squares} (\qcels{}) algorithm \cite{PRXQuantum.4.020331} to the problem. We investigated the suitability and limitations of the \qcels{} algorithm when applied to large systems \cite{QCELS_for_QAGC}. In the evaluation, we have assessed their performance and potential benefits for solving 16 quantum systems of 20-, 24- and 28-qubits, showing improvement for some of the 20- and 28-qubit systems.

The rest of the paper is structured as follows. \sectref{sec:RW} lists related work in the fields of quantum computing and machine learning for quantum. \sectref{sec:back} focuses on the background of the quantum computing algorithms that we have investigated, software engineering optimisation methods and machine learning, followed by a deeper dive into the methodologies of quantum eigensolver algorithms in \sectref{sec:solvers}, and machine learning applications to accelerate quantum algorithms in \sectref{sec:preml}. In \sectref{sec:results}, we present the experiment's methodology and our results. Our conclusions can be found in \sectref{sec:conc}, including notes containing references to the source code used in the paper and the relevant data and acknowledgements.

\paragraph{A note on the 2024 Quantum Algorithm Grand Challenge (QAGC2024). } The challenge focused on improving current quantum algorithms for the problem of finding the ground state energy of a one-dimensional orbital rotated Fermi-Hubbard model Hamiltonian using a given 28-qubit system in the most optimal way \cite{QAGC2024}. The aim of the challenge is to design an algorithm, given specific constraints (\textit{i.e.} the number of shots of $10^7$, timeout of $6 \times 10^5$ seconds and limited input data), that can output the closest answer to the actual ground state energy. The 2023 challenge \cite{QAGC2023} had the same aim but was based on a much smaller 8-qubit system. The winning algorithm was the company's proposition, later published as Quantum-Selected Configuration Interaction (QSCI)~\cite{QSCI2023}. 
\section{Related Work}
\label{sec:RW}

\noindent\textbf{Challenges in Quantum Computing.}
Quantum Computing (QC) was conceptualised initially to simulate quantum mechanics using computers \textit{``built of quantum mechanical elements which obey quantum mechanical law''} \cite{feynman_simulating_nodate}. Later, it was found that QC could have several potential applications and offer significant speed-up over classical computing \cite{QCrypto1983,Qkeydist1987,deutsch1992rapid,PhysRevLett.70.1895Qteleport1993,grover_fast_1996,Brassard2016}.
In 1994, Shor's proposal of a polynomial-time algorithm for prime factorization and discrete logarithms on a quantum computer raised enormous interest due to its potential threat to modern RSA cryptosystems \cite{shor_polynomial-time_1997}. Soon after, Grover introduced a fast database search on quantum computers \cite{grover_fast_1996} that promised quadratic speed-up over the best classical algorithm. The resulting potential speed-up is often referred to as ``quantum supremacy'' \cite{arute_quantum_2019}. Several studies apply software engineering techniques to optimise quantum computing \cite{greiwe_effects_2023, gheorghe-pop_quantum_2020}. Noticeably, testing \cite{miranskyy_testing_2019, supic_self-testing_2020}, debugging \cite{paltenghi_bugs_2022, sato_locating_2023, metwalli_tool_2022}, verification \cite{yan_incorrectness_2022, li_verified_2022} approaches, and efficient synthesis techniques \cite{kang_modular_2023, venev_modular_2024, paradis_synthetiq_2024} have been found to be beneficial in quantum software development \cite{miranskyy_testing_2019, supic_self-testing_2020, paltenghi_bugs_2022, yan_incorrectness_2022}. 

Demonstrating quantum supremacy on real hardware remains a long-standing challenge, especially at a scale where quantum devices would solve real-life calculations. Although quantum supremacy seems difficult to achieve soon, NISQ algorithms---Imperfect hardware is often called Noisy Intermediate-Scale Quantum (NISQ) devices---are a prominent example that hybrid systems combining small quantum circuits with classical computations could present some computational advantages, i.e., a quantum advantage \cite{preskill_quantum_2018}. Most agree this stage of QC will likely last for the next few years if not decades, and refer to it as the NISQ era \cite{preskill_quantum_2018}. Variational Quantum Algorithms (VQA) are the most common example of an efficient combination of a reduced quantum circuit inside a classical optimisation loop \cite{tilly_variational_2022}. Other algorithms use classical optimisation to optimise quantum calculations, such as QCELS \cite{ding_even_2023, ding_simultaneous_2023} that uses a fitting procedure to optimise quantum calculations. Because of their prominent role in modern Quantum Computing research and industrial applications, we chose to focus our study on Variational and QCELS Algorithms.

\vspace{0.2 cm}
\noindent\textbf{Machine Learning for Quantum Software Engineering.}
Machine learning algorithms are increasingly used to improve and automate software engineering tasks \cite{ML2SE1997,ML2SE2003,ML2SE_Harman2012,ML4SE2018,ML4SE2019}, especially after the advent of Large Language Models (LLM), 
with common applications in software engineering, including optimisation, code generation, bug detection and automated testing \cite{10449667,wang2024software,GinLLM,StableYolo2023,Dakhama2023:searchgem5,huang2024large}. In \autoref{sec:preml}, we suggest applying these ideas in the context of quantum algorithms, accelerating their performances 
We apply a search-based process to identify the optimal combination of the hyperparameters (global minima) using our regressor trained on smaller systems to estimate the system energy as the fitness function.  
%%%
We give here an overview of the methods used: search-based methods, the regressor and its Python implementation. \cite{fastovets_machine_2019, wang_comprehensive_2024, peral-garcia_systematic_2024}.

Although existing applications of Machine Learning in Quantum Computing have been investigated \cite{ramezani_machine_2020}, it remains a very young and open field of research. While Quantum Machine Learning offers potential speedups in Machine Learning tasks \cite{umer_comprehensive_2022}, very recent work also demonstrates promising results in applying Machine Learning to Optimise Quantum Computations \cite{cho_machine_2024, li_quarl_2024}.

\section{Background}
\label{sec:back}

Our approach combines machine learning, search-based software engineering, and quantum physics to accelerate quantum algorithms, specifically eigensolvers. We provide background on each area before discussing the applications of machine learning in quantum algorithms. In \sectref{sec:solvers}, we discuss the implementation of the eigensolvers used in this research.

\subsection{Variational Quantum Algorithms}
\label{sec:vqa}
Variational Quantum Algorithms (VQA) are among the most promising examples of NISQ algorithms \cite{tilly_variational_2022}. 
The main goal of a VQA is to find the optimal parameters for a parameterized quantum circuit, leading to a solution for a given computational problem. We provide an overview of the structure and functioning of a VQA. 

\begin{enumerate}

\item  \textit{Problem definition.} The first step involves defining a computational problem that can benefit from quantum processing and translating it into an objective or cost function. In most applications, it consists of a Hamiltonian construction and representation of a quantum system.

\item  \textit{Optimisation process.} A parameterized quantum circuit, known as the variational ansatz, is designed. This circuit contains gates with adjustable parameters, denoted as $\theta$. A quantum state $\ket{\psi(\theta)}$ is measured, and the outcomes are used to compute the expectation value of the cost function $\bra{\psi(\theta)}H\ket{\psi(\theta)}$.
H is the Hamiltonian of the quantum system, and the expectation value of this Hamiltonian is the cost function.

\item  \textit{Convergence check and output.} The optimisation process continues until a convergence criterion is met, indicating that further iterations are unlikely to significantly improve the solution. This convergence check ensures that the algorithm has reached a stable and potentially optimal solution. The final set of optimised parameters $\theta_{opt}$ represents the solution to the quantum problem. This solution can be used for further analysis or as the output of the VQA.

\end{enumerate}

An in-depth explanation and a detailed figure can be found in the work of Bharti et al. \cite{bharti_noisy_2022}. 

\paragraph{VQE and ADAPT-VQE. } Variational Quantum Eigensolvers (VQE) were introduced in 2014 by Peruzzo et al. \cite{peruzzo_variational_2014}. It is the most popular VQA, and aims to approximate the ground state energy of a quantum system iteratively. Although considered among the most promising NISQ algorithms, VQEs present several limitations. VQE optimisations have been widely investigated, from classical optimisers, to measurement strategies and Ansatz structure \cite{tilly_variational_2022}. One popular algorithm derived from VQEs is the ADAPT-VQE \cite{grimsley_adaptive_2019}. The main distinction is the restructuring of the Ansatz at each iteration. A pool of operators is defined from which operators are selected to update the ansatz during the optimisation process. ADAPT-VQEs are found to have more precise results in some applications, at a cost of a higher computational load.

%% Future work: try Bayesian Optimization: Efficient for complex or high-dimensional spaces, often providing faster convergence to optimal hyperparameters. KE: not better, but reviewers will ask.
\subsection{Machine Learning and Software Engineering Methods}
\label{sec:background:SEML}

In this work, we integrate ML with search-based methods. We first use predictions on smaller systems, utilising the cost function values of a Hamiltonian of smaller systems and randomly generated hyperparameters of the quantum algorithm to train a gradient boosting model to predict the cost function value. We then use this model to explore and optimise hyperparameters for a larger Hamiltonian system.
%This stage of the optimisation is performed by applying Genetic Improvement (GI) using a crossover operator, with predictions of the quantum system’s cost function guiding the selection process. This approach integrates Machine Learning (ML) for predictive modelling with search-based optimisation to explore solutions, and then Genetic Improvement (GI) for evolutionary optimisation for the larger system. We discuss each of the approaches.

\paragraph{Search-based Software Engineering. }
Search-based methods \cite{HARMAN2001833} have also been used for optimisation tasks to find the best possible subset of requirements. 
The optimisation, in the case of the quantum algorithm's hyperparameters, aims to save precious resources like the number of shots but mainly to achieve a closer prediction to the system's ground state energy\footnote{The lowest energy level, which for eigensolvers is the lowest eigenvalue that is the value associated with the lowest eigenvector of the problem, see \sectref{sec:solvers}.}, influenced by the selection of optimal hyperparameters (i.e. a minimum of the cost function approximating the lowest energy level, \subsectref{sec:vqa}).

These methods aim to find the near-optimal solution iteratively, ending when a stopping condition is met (e.g. reaching a maximum number of iterations).
Starting with a set of candidate solutions, they are evaluated using a fitness function. 
In this work, we utilise the \textit{genetic algorithms} (GA) approach. GA diversifies solutions over iterations through mutation operations like bit flip (on a single solution) or crossover (on multiple solutions) to potentially generate better-performing offspring. The method can occasionally minimise the population or inject noise to avoid converging to local minima (or maxima, depending on the problem). Using ML to estimate the fitness function is common when direct evaluation or computation is infeasible. We utilise this idea for larger Hamiltonians in \sectref{sec:preml}.

In the context of this work, we define a solution as an assignment of hyperparameters (their values based on some pre-defined constraints, e.g. \code{sampling\_shots} are between $100$ to $10^6$), with the best solution being a global minimum.
Consequently, our fitness function is related to the lowest energy level for a specific assignment of the hyperparameters and Hamiltonian.
We perform these stages classically as quantum computations are costly. However, the final estimation of the quantum algorithm's cost function is done using a quantum simulator with the optimised hyperparameters.

\paragraph{Machine Learning. }
ML, a branch of artificial intelligence (AI), aims to learn from data and generalise models via statistical algorithms (e.g. random forest or linear regression) to identify patterns and make predictions and decisions on new, unseen data. The process typically involves several main phases: data collection, data pre-processing (including data augmentation and mining), model training, and model prediction (or deployment). 

In this work: Data consists of Hamiltonians, parameters and the lowest energy levels. We collect data from smaller quantum systems (16 qubits and below) to build a model for larger systems (20 qubits and above). Data is gathered from online sources or computed via simulations or classical algorithms. The data is then processed and augmented to ensure compatibility with machine learning models, aiming to keep relevant features of the physical system for training. We discuss our choices with respect to data collection, augmentation, and model prediction in \sectref{sec:preml}.

\paragraph{Gradient Boosting. } The Gradient Boosting technique \cite{friedman2001greedy} is a supervised learning method suitable for regression and classification. Supervised machine learning utilises labelled data to train predictive models \cite{kotsiantis2007supervised}.

Gradient boosting efficiently handles large amounts of data, real numbers, and datasets with many features, as is the case here when representing the Hamiltonian in our dataset, which, even with aggressive approximation, may result in large data instances. The resultant prediction model allows
%is in the form of an ensemble of weak prediction models (usually decision trees), allowing 
prediction with large datasets with high precision, avoiding flat predictions caused by scaling issues or oversimplified predictions due to insufficient data relative to the number of features.

\paragraph{XGBoost Library. } We employ XGBoost (eXtreme Gradient Boosting), an open-source Python library of the Gradient Boosting framework \cite{xgboost}, as a regressor to predict a response variable, in our case, the lowest energy level of the quantum system (the Hamiltonian). The input features for our model include the Hamiltonian, the number of qubits, and a set of hyperparameters.
\section{Quantum Eigensolver Algorithms}
\label{sec:solvers}

During our experiments and development of the solution to the challenge, we used and wrote two implementations of \textit{Quantum Eigensolver} (QE), \adapt{} and QCELS, described in \subsectref{VQE:alg} and \subsectref{method:qcels}.
QEs aim to find the nearest approximation of the eigenvalues and eigenvectors of a given system (i.e. the input Hamiltonian). Furthermore, in QC, we aim to determine the ground state energy of the system (the lowest energy level), similarly to the challenge \cite{QAGC2024}. 

In our evaluation, we used the implementation of \adapt{} provided by the challenge scripts \cite{QAGC2024} and implemented our own \qcels{} solver. Each implementation can run on a classical statevector (exact) simulator (i.e. classic mode) or on a matrix product state (MPS) simulator (i.e. QC mode) \cite{PhysRevX.10.041038,PRXQuantum.4.020304,PhysRevA.106.052430,Noh2020efficientclassical}, provided as part of the challenge, which is suitable for simulating large system sizes by enabling a trade-off between speed and accuracy. In classical mode, we are limited by the size of the system and probably can run up to 20-qubit systems at most. In our experiments, we limited our classical mode runs to 16-qubit systems. 

Next, we describe the two QE algorithms used in this work.
%\all{Add a note on being able to run it on classical simulator.}
%\all{does it have a classical version? Yes we can run it classically as well, probably the system sizes this would be achievable for would be smaller. KEM: can we add such a note? What is the expected/known size we can work on? We probably need to add it to 3.1 too.}

\subsection{\adapt{} Algorithm}
\label{VQE:alg}

\paragraph{QSCI.} 

Quantum-selected configuration interaction (QSCI) method is a computational algorithm in quantum chemistry that is used for calculating electronic structure of molecules in an intelligently chosen subspace that makes larger systems feasible to study on modern-day NISQ devices \cite{QSCI2023}. 
A full configuration interaction requires high computational cost and memory usage that is out of reach for large systems but by using QSCI the computational space can be reduced through selecting only the most important configurations (ways of distributing electrons among molecular orbitals) by a pre-selection algorithm. One such algorithm is \adapt{} \cite{ADAPTQSCI2023}.

\paragraph{\adapt{}.} 

Adaptive Construction of Input State for Quantum-Selected
Configuration Interaction (\adapt{}) is an iterative algorithm that uses a predetermined pool of single Pauli operators $\probP = \{P_1,\dots , P_T \}$ that are generators of rotation
gates for the input quantum state of QSCI, and selects the best operators from the pool to lower the energy output by QSCI. The operator pools are similar to ADAPT-VQE approaches and could be based on fermionic or qubit excitations. In this method a simple sampling measurement is performed on an input state prepared by a quantum computer - this is the only step that requires quantum computation. The measurement result is used for identifying the most important electron configurations for performing the selected configuration interaction calculation on classical computers, that is, Hamiltonian diagonalization in the selected $R_k$ dimensional subspace $S_k = \operatorname{span}\{\ket{r_1^{(k)}}, \ldots, \ket{r_{R_k}^{(k)}}\}$ \cite{ADAPTQSCI2023}. QSCI relies on a quantum computer only for generating the electron configurations via sampling, and the subsequent calculations to output the ground-state energy and the pool operator gradients $h_j = \braket{c_k|i[H, P_j ]|c_k}$ are executed on classical computers\footnote{$\ket{c_k}$ is a state corresponding to classical vector $c_k=\sum_{l=1}^{R_k} (c_k)_l\ket{r_l^{(k)}}$, and $i[H, P_j ]$ is calculated through projecting onto the subspace $S_k$ and evaluating the expectation classicaly using the classical vector $c_k$.\cite{ADAPTQSCI2023}.}. These calculations are made possible on classical machines due to the reduction of the dimensionality of the subspace based on QSCI. The quantum advantage of the \adapt{} stems from the potential speed up and improved precision of the generating of the electron configurations via sampling.

\subsection{Quantum Complex Exponential Least Squares (\qcels{}) Algorithm} 
\label{method:qcels}

To go beyond NISQ algorithms, where typically the QC is used to simply prepare an ansatz state and then measured in a Pauli basis, we used the Quantum Complex Exponential Least Squares algorithm (\qcels{}) \cite{PRXQuantum.4.020331} which uses a controlled time evolution unitary and so introduces a large number of two-qubit gates. However, it does not suffer from the barren plateau problem of VQE and is low enough depth that it will likely be run on early QCs with some error correction, offering a good solution when the number of qubits is too large for NISQ algorithms.

\begin{figure}[ht!]
\vspace{0.4cm}
\centering
\resizebox{!}{0.364cm}{
\Qcircuit @C=1em @R=.7em {
&&&  \lstick{\ket{0}}    & \gate{H} & \ctrl{1} & \gate{W} & \gate{H} & \meter & \qw \\
&&&  \lstick{\ket{\psi_0}} & \qw & \gate{e^{-it_{n}H}} & \qw & \qw & \qw & \qw \\
}
}
\caption{\qcels{} circuit \cite{PRXQuantum.4.020331}.}
\label{fig:qcels}
\end{figure}
%% https://mirrors.ibiblio.org/CTAN/graphics/pgf/contrib/quantikz/quantikz.pdf

The \qcels{} algorithm \cite{PRXQuantum.4.020331} takes a reference state \(\ket{\psi_0}\) and evolves it by the time evolution operator \(U(t) = e^{-iHt}\) enclosed within a Hadamard test (as depicted by Figure~\ref{fig:qcels}). If the reference state is the ground state (\(\ket{\phi_0}\)) then the resultant expectation values will have a single frequency  \(Z_n = \braket{\phi_0|U(t_n)|\phi_0} = e^{-iE_0 t_n}\), where $E_0$ is the ground state energy. If the reference state, $\ket{\psi_0}$, is not exactly the ground state then the resultant function is \(\ Z_n \approx \braket{\psi_0|U(t_n)|\psi_0} = \sum_i p_i e^{-iE_i t_n}\)  where \(p_i = |\braket{\phi_i|\psi_0}|^2
\) is the probability of measuring the eigenstate \(\ket{\phi_i}\) of the Hamiltonian. Thus, if a reference state with good overlap with the true ground state is known, we can apply the time evolution operator within a Hadamard test $N$ times and fit to the resulting complex exponential. 
For this challenge, we found that the provided Hamiltonians had an initial  Hartree-Fock ground state that retains an overlap with the true ground state - up to \(28\) qubits - to extract a good estimate of the energy. The time evolution operator was implemented with a first-order Trotter-Suzuki expansion with Hamiltonian truncated to only the Pauli operators with the largest coefficients to reduce gate count.

Our procedure for fitting a sum of exponentials to the collected data was to initially fit with a single frequency fitting to 
\begin{equation}
    f_{\text{fit}}^{(1)} = r_1^{(1)} e^{-i \theta_1^{(1)} t} + 1 - r_1^{(1)},
\end{equation}
before using this frequency as an initial guess and then adding sequentially second and third frequencies,
\begin{align}
    f_{\text{fit}}^{(2)} &= r_1^{(2)} e^{-i \theta_1^{(2)} t} + r_2^{(2)}e^{-i \theta_2^{(2)} t} + 1 - r_1^{(2)} -r_2^{(2)} \\
    f_{\text{fit}}^{(3)} &= r_1^{(3)} e^{-i \theta_1^{(3)} t} + r_2^{(3)}e^{-i \theta_2^{(3)} t} + (1 - r_1^{(3)} -r_2^{(3)})e^{-i \theta_3^{(3)} t}
\end{align}
which were constrained to be smaller in magnitude than the initial fitted frequency $\theta_1^{(1)}$ to ensure that they acted as corrections to the initial fit, for the purpose of the challenge we also added a factor to keep the change in $\theta_1$ small ensuring that the fitting procedure is as reliable as possible due to the need for it to be automated. The size of the amplitude $r_2^{(2,3)}$ is also constrained to be smaller than $r_1^{(1)}$.

\begin{figure}[ht]
    \centering
    \begin{subfigure}[b]{0.32\linewidth}
        \centering
        \includegraphics[width=\linewidth]{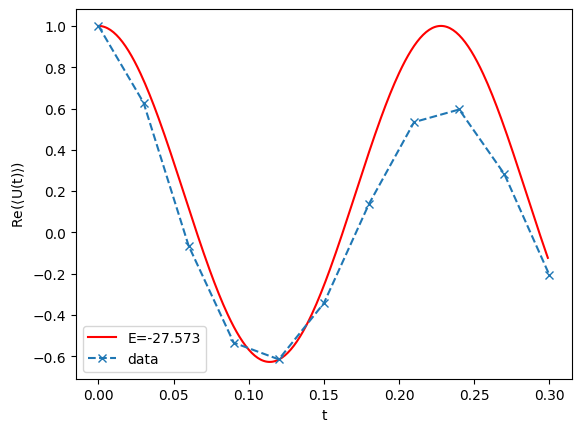}
        \caption{$f_{\text{fit}}^{(1)}$}
        \label{fig:28qbs_1_freq}
    \end{subfigure}
    \hfill
    \begin{subfigure}[b]{0.32\linewidth}
        \centering
        \includegraphics[width=\linewidth]{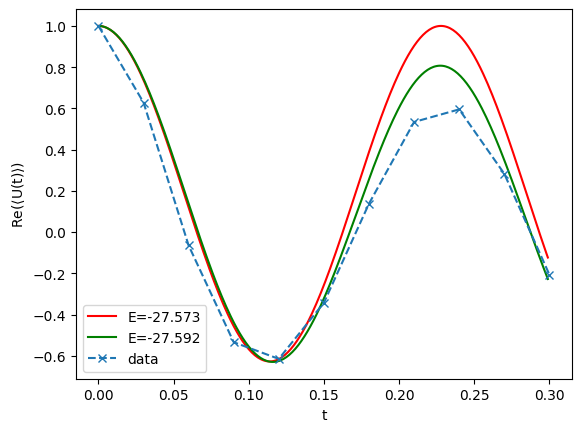}
        \caption{$f_{\text{fit}}^{(2)}$}
        \label{fig:28qbs_2_freq}
    \end{subfigure}
    \hfill
    \begin{subfigure}[b]{0.32\linewidth}
        \centering
        \includegraphics[width=\linewidth]{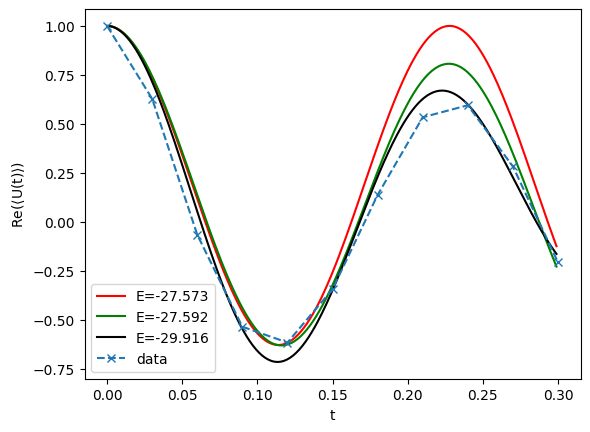}
        \caption{$f_{\text{fit}}^{(3)}$}
        \label{fig:28qbs_3_freq}
    \end{subfigure}
    \caption{\qcels{} results from sample data of the orbital rotated Hubbard Hamiltonian on 28 qubits (various parameters).}
    \label{fig:28qbs_fitting}
\end{figure}

The outcome of this procedure when applied to test data collected from an example of the orbital rotated Hubbard Hamiltonian on 28 qubits is given in figure \autoref{fig:28qbs_fitting}, with a sequential improvement towards the correct answer as we increase the number of frequencies in the fit. 
%refining our guess by allowing for more energies to contribute.  
%Use a hierarchical fitting procedure

%%  https://github.com/QunaSys/quantum-algorithm-grand-challenge-2024/blob/main/problem/evaluator.py

% \begin{figure}[ht]
%     \centering
%     \begin{subfigure}[b]{0.49\linewidth}
%         \centering
%         \includegraphics[width=\linewidth]{Figures/12qbs-collect.png}
%         \caption{Caption for first figure}
%         \label{fig:fig1}
%     \end{subfigure}
%     \hfill
%     \begin{subfigure}[b]{0.49\linewidth}
%         \centering
%         \includegraphics[width=\linewidth]{Figures/12qbs-fit.png}
%         \caption{Caption for second figure}
%         \label{fig:fig2}
%     \end{subfigure}
% \end{figure}

\section{Accelerating Quantum Algorithms with Machine Learning}\label{sec:preml}

%\all{It is worth clarifying: model, system and problem in the context of this work, which can be a bit confusing.}

In the previous section, we explored two implementations of the QE algorithms: \adapt{} and \qcels{}. Building on these foundations, we now introduce our proposed solution, the prototype of which was submitted to the QAGC challenge \cite{QAGC2024}. 
Our approach enhances \adapt{} and \qcels{} by integrating machine learning techniques to optimise their hyperparameters. 
Before presenting our solution, we explain how we gather sufficient data to train an ML model (\subsectref{sec:classicalwrapper}). 
We then present our solution in detail (\subsectref{sec:tool}), followed by implementation details and the constraints set by the QAGC challenge rules (\subsectref{sec:qagc}).
In the evaluation, we demonstrate that quantum algorithms can benefit from incorporating a machine learning model to improve decision-making and overall performance (\sectref{sec:results}).

%%%%%%%%%%%%%%%%%%%%%%%%%%%%%%%%%%%%%%%%%%%%%%%%%%%%%%
%%%%%%%%%%%%%%%%%%%%%%%%%%%%%%%%%%%%%%%%%%%%%%%%%%%%%%
%%%%%%%%%%%%%%%%%%%%%%%%%%%%%%%%%%%%%%%%%%%%%%%%%%%%%%
%%%%%%%%%%%%%%%%%%%%%%%%%%%%%%%%%%%%%%%%%%%%%%%%%%%%%%
\subsection{Data Augmentation}
\label{sec:classicalwrapper}

ML algorithms operate in two primary phases: training and prediction. During the training phase, a sufficiently large dataset is essential for making probabilistic generalisations. In a supervised learning context, this dataset comprises instances where the true output values are known. Once the model is trained, it can make predictions on new, unseen data. The quality and accuracy of the training dataset directly influence the effectiveness of these predictions. Consequently, our initial challenge was to obtain a robust initial dataset.

Acquiring a sufficiently large dataset is a significant challenge. One common solution to this problem is \emph{data augmentation}, a technique that involves creating additional training data from the existing dataset through various transformations. These transformations can include rotations, scaling, cropping, and other modifications that simulate new data samples, thereby enhancing the diversity and volume of the training dataset. The effectiveness of data augmentation has been extensively studied and validated in numerous literature reviews (\emph{e.g.,}~\cite{maharana2022review, mikolajczyk2018data, wang2017effectiveness, shorten2019survey, wong2016understanding}).

The accuracy and generalisation of an ML model are closely linked to the size of the dataset. In the context of quantum simulation data, each data instance encompasses numerous features due to the substantial memory required to represent the Hamiltonian. To generate a sufficiently large dataset for training, we utilise a classical state vector simulator rather than a quantum device, as quantum computing time is prohibitively expensive. This simulator allows us to compute the exact energy levels of smaller Hamiltonians (\emph{i.e.} fewer than 28 qubits) with high precision. Each data instance in our training dataset encompasses the complete set of information required for the quantum problem, including the Hamiltonian\footnote{In practice, we do not include in our data the Hamiltonian as-is but a compressed representation of the Hamiltonian to save memory and avoid overfitting. Compressing a Hamiltonian in the context of QC reduces the complexity of the Hamiltonian operator but keeps the essential physical properties of the system. We compressed by truncating the smallest terms}, its exact energy level, and the hyperparameters used to compute this level.

We use this dataset to train our ML model and predict hyperparameters for new Hamiltonians. We anticipate that the generalisation capabilities of deep learning will enable our model to handle larger Hamiltonians and predict hyperparameters for unseen 28-qubit problems effectively. This approach adheres to the challenge rules outlined in \subsectref{sec:qagc}, as we do not apply classical methods to Hamiltonians of 28 qubits or larger. Further details on the dataset structure and mining processes are discussed in the \emph{Data Preparation} stage described in \subsectref{sec:tool}.

% a note on regressor.
Note on the energy levels predictions: As energy levels are real numbers (i.e., are in a continuous domain), classification might result in precision issues during the prediction phase. Moreover, the exact energy level value is irrelevant as we do not use these values directly. We investigate whether specific data (Hamiltonian and quantum-system-related parameters) contribute to further minimising the lowest energy level. We are interested in exploring the relationship between these values. 
Therefore, we have used a regressor to predict roughly the lowest energy levels. %We discuss the specific technique and implementation used in this work below.

%%%%%%%%%%%%%%%%%%%%%%%%%%%%%%%%%%%%%%%%%%%%%%%%%%%%%%
%%%%%%%%%%%%%%%%%%%%%%%%%%%%%%%%%%%%%%%%%%%%%%%%%%%%%%
%%%%%%%%%%%%%%%%%%%%%%%%%%%%%%%%%%%%%%%%%%%%%%%%%%%%%%
%%%%%%%%%%%%%%%%%%%%%%%%%%%%%%%%%%%%%%%%%%%%%%%%%%%%%%
\begin{figure}[t!]
    \centering
    \includegraphics[width=0.95\linewidth]{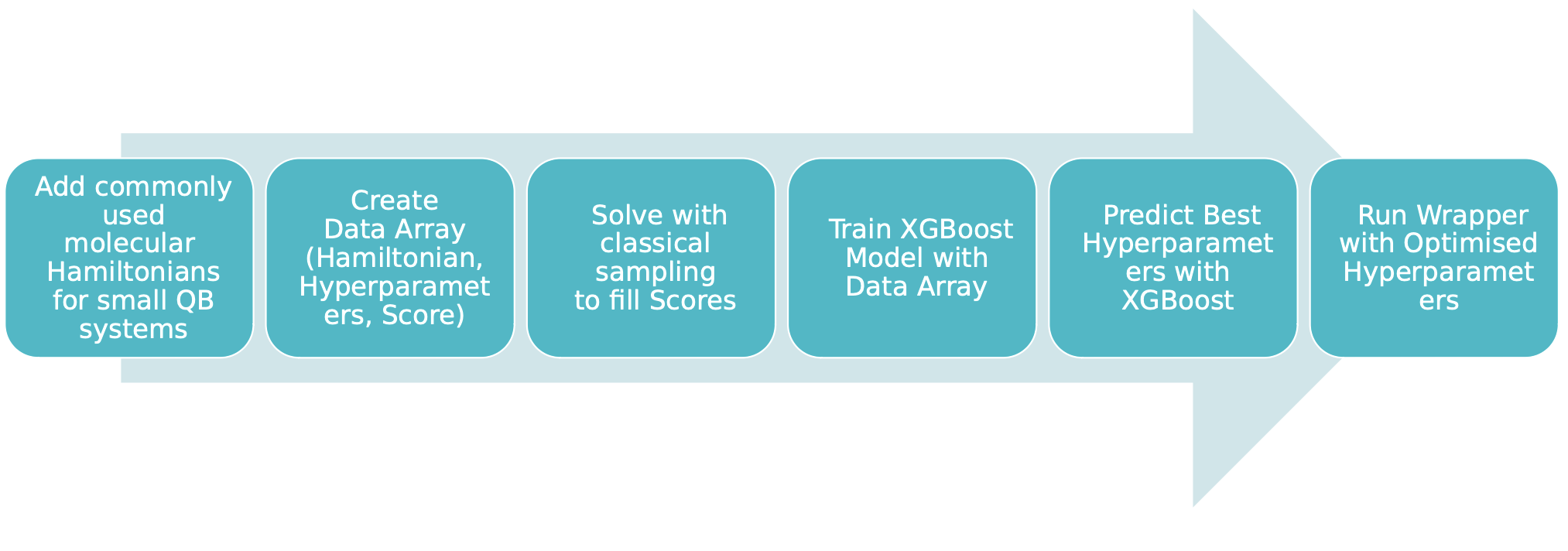}
    \vspace{-0.3 cm}
    \caption{Classical pre-processing phase - training a regressor on smaller systems.}
    \label{fig:hyprparam:opt:flow}
\end{figure}

\subsection{AccelerQ}
\label{sec:tool}

In this section, we present \tool, our approach for enhancing quantum algorithm performance through hyperparameter optimisation. We employ a search-based methodology to identify the optimal combination of hyperparameters by leveraging a regressor trained on smaller systems. This regressor estimates system energy, which serves as the fitness function in our optimisation process. Below, we provide an overview of the methods employed, including search-based techniques, the regressor, and its Python implementation.

\autoref{fig:hyprparam:opt:flow} illustrates our hyperparameter optimisation algorithm for quantum problems, which utilises a regularizing gradient boosting regressor (XGBoost). The process is organised into three key stages: data preparation, model training, and hyperparameter optimisation. This algorithm generates tailored hyperparameter suggestions for executing the QE algorithm, customised for each specific Hamiltonian.

To ensure the generalisability of \tool and its applicability to various QE algorithms (e.g., \adapt{} or \qcels{}), we have developed a versatile wrapper. This wrapper encompasses the Hamiltonian, the number of qubits, a parameter indicating whether the computation should be performed classically or on the challenge MPS simulator, and the hyperparameters (of both, the QE and ML problems, while we only optimise here those of the QE problem). In our evaluation (see \sectref{sec:results}), we applied \tool to the \adapt{} and the \qcels{} algorithms.

\paragraph{Data Preparation. }
We use Hamiltonians from \cite{QAGC2024} and open-source commonly used molecular Hamiltonians (H2O, LiH, BeH2, Hemocyanin \cite{Cedric-hemocyanin}, and Hydrogen chain) of \(16\) qubits. The wrapper is run in classical mode while giving a randomly generated set of hyperparameters each time, recording the energy level as the score. This process is repeated to produce a data array for training a Regularising Gradient Boosting Regression (XGBoost). The generated data—consisting of a compressed Hamiltonian and hyperparameter vectors (Xs) and their corresponding energy levels (Ys)—is stored for further processing. This stage corresponds to the first of the three steps depicted in~\autoref{fig:hyprparam:opt:flow}.
We utilised a quantum eigensolver run on an emulator during data preparation (\subsectref{sec:classicalwrapper}).

\paragraph{Model Training. }
Our algorithm uses an XGBoost model to predict the best hyperparameters. Before training, it ensures that all data vectors are of consistent length by padding them as needed. The padded data array is split into training and testing sets. The model is then trained, tested, and evaluated for performance. The training of the XGBoost model corresponds to the fourth step of~\autoref{fig:hyprparam:opt:flow}.

\paragraph{Hyperparameter Optimisation. }
With the XGBoost model trained, the algorithm proceeds to predict the optimal hyperparameters for a given Hamiltonian of \(28\) qubits (\autoref{fig:hyprparam:opt:flow}, fifth step). It generates a series of hyperparameter vectors and uses the XGBoost model to predict their performance. 
In each iteration (generation), the best-performing vectors are selected (i.e. predicted to have minimum score value). A crossover operator combines pairs of these vectors, generating new hyperparameters through averaging, extremes, and random values. 
In the last generation, the vector with the minimum score (predicted energy level) is returned as the optimal set of hyperparameters. Finally, the algorithm runs \adapt{} with the optimised hyperparameters (\autoref{fig:hyprparam:opt:flow}, sixth and last step).

%% TODO: add in the final version
%%\subsection{Generalisation}
%%\all{Add text from ReadMe}
%%General wrapper and generator, to do 4.1 as we wish.

%%%%%%%%%%%%%%%%%%%%%%%%%%%%%%%%%%%%%%%%%%%%%%%%%%%%%%
%%%%%%%%%%%%%%%%%%%%%%%%%%%%%%%%%%%%%%%%%%%%%%%%%%%%%%
%%%%%%%%%%%%%%%%%%%%%%%%%%%%%%%%%%%%%%%%%%%%%%%%%%%%%%
%%%%%%%%%%%%%%%%%%%%%%%%%%%%%%%%%%%%%%%%%%%%%%%%%%%%%%
%\all{This probably should be part of implementation details as it dictated python and packages versions and other limitations. - For paper version} 
\subsection{Implementation Details}
\label{sec:qagc} 

Our implementation is divided into three parts (data collection, training and optimisation) and is written in Python 3.10.12. The rest of the requirements are derived from the rules of the 2024 Quantum Algorithm Grand Challenge (QAGC2024), with one exception: we utilised GPU during the training phase of our models while keeping them relatively small to be able to deploy them on desktop CPU. We used the Python library \textit{QURI Parts}\footnote{\href{https://pypi.org/project/quri-parts/}{https://pypi.org/project/quri-parts/}} to simulate the QC.

\paragraph{Quantum Algorithm Grand Challenge Description. }
Participants in the 2024 Quantum Algorithm Grand Challenge (QAGC2024) were tasked with challenging the \adapt{} Algorithm~\cite{ADAPTQSCI2023}, a quantum-classical hybrid approach designed to compute the ground states and energies of many-body quantum Hamiltonians (\subsectref{sec:vqa}). The challenge specifically focuses on optimising a quantum algorithm for solving a \(28\)-qubit system under several constraints (see full list~\cite{restrictions2024}). These constraints included a specific machine specification (CPU, 16 GB RAM), a practical time limit of \(6 \times 10^5\) seconds to ensure all solutions could be evaluated within a reasonable timeframe and a shots limit of \(10^7\) due to the high cost of quantum computer execution. Participants were required to make essential use of quantum computers, as the goal was a quantum competition, and current classical algorithms do not scale efficiently for systems with many ($\gtrsim 50$) qubits.

Furthermore, participants were instructed to output only the energy level of the ground state, calculated using the quantum algorithm of choice. Exact calculations of the ground-state energy using methods like the Bethe ansatz were prohibited, as such methods are either inefficient or do not work for general systems. Additionally, participants were not allowed to hard-code initial parameters or hyperparameters to ensure adaptability to new Hamiltonians. That constraint ensures that the proposed solution is generic and not tailored for a specific Hamiltonian or subset of problems.

To facilitate the automatic evaluation of all participants' solutions, the contest organisers provided a specific structure for their responses, limiting participants to modifying a single \texttt{answer} file. While this constraint streamlined the evaluation process, it also introduced certain limitations. Notably, it prevented us from leveraging the full potential of machine learning techniques, as models could not be stored globally and had to be retrained for each evaluation. Although this approach was effective for the contest evaluation, it is not ideal for software engineering practices that prioritise modularity. A more effective strategy would be to divide the problem into distinct stages, each with its own constraints. Constraints provided by the contest organisers are particularly relevant for the prediction phase but may not be suitable for the training phase. Recognising these limitations, in future iterations of our solution, we plan to adopt a more flexible approach, allowing for greater use of advanced methods and better adaptability while respecting the original framework and intentions set by the challenge organisers.

\section{Evaluation}
\label{sec:results}

We evaluated a prototype of \tool{} for its ability to optimise further \adapt{} and \qcels{} algorithms by suggesting better hyperparameters tailored per system, in this report focusing on random search without a crossover operator.

\subsection{Methodology}
\label{sec:methodology}

We demonstrated our idea by optimising hyperparameters of two quantum eigensolver algorithms. We used 4- to 16-qubit systems to train the models. We deployed the models on 20-, 24- and 28-qubit systems. We aim to understand:
\begin{quote}
\textbf{RQ1:} To what extent can \tool's optimisation of hyperparameters alone accelerate and improve the efficiency and accuracy of QE algorithms on NISQ devices in terms of %runtime, 
error, and system size?
\end{quote}

Given that we extract a model from smaller systems up to 16 qubits, we wish to assess the scalability of these models and to test if we have reduced performance as the system gets larger, that is:
\begin{quote}
\textbf{RQ2:} 
How scalable are machine learning-predicted hyperparameters learned on smaller systems when applied to QE algorithms for Hamiltonian systems with increasing qubit number and complexity?
\end{quote}
In our evaluation, we applied our methodology to 16 larger systems of 20-, 24- and 28-qubit with known lowest energy levels\footnote{In the case of the open-source systems, we have only a close estimation using \adapt{} hyperparameters results as a baseline.}, using the two QE implementations: \adapt{} and \qcels{} (\sectref{sec:solvers}) to answer RQ1 and RQ2. 

We constructed two models—one for each implementation—using data extracted classically (\sectref{sec:solvers}) from up to 16-qubit systems. The data extraction and model training were general. The resultant model aimed to predict the optimal hyperparameters for its QE implementation and a system (Hamiltonian). The optimisation process took an implementation, its trained model and a system (a problem to solve) as input, and returned the predicted optimal hyperparameters for this setup. 

\paragraph{Baseline and Parametrisation. } Our evaluation baseline is the results obtained with an implementation's default hyperparameters, fixed across all systems in the evaluation. The hyperparameters were specific per implementation, were optimised, and their performance was compared against the baseline. Apart from the hyperparameters, \adapt{} and \qcels{} implementations were provided with 
\textbf{ham} and \textbf{number\_qubits}, the input system and the required number of qubits for the corresponding Hamiltonian, and the flag \textbf{is\_classical} being set to \code{True} during the data collection phase, and either \code{True} or \code{False} otherwise (exact classical simulation or MPS). We did not optimise this part.

%%
%%        self.number_qubits = number_qubits
%%        self.ham = ham
%%        self.is_classical = is_classical #use SV solver
%%        self.use_singles = use_singles #include single excitations in the operator pool
%%        self.num_pickup = num_pickup #retain largest N terms in Hamiltonian
%%        self.coeff_cutoff = coeff_cutoff #cutoff all terms smaller than this from the num_pickup terms remaining
%%        self.iter_max = iter_max #max total iterations
%%        self.vqe_shots = vqe_shots #how many shots to use per iteration
%%        self.grad_shots1 = grad_shots1
%%        self.grad_shots2 = grad_shots2
%%        self.precise_grad = precise_grad
%%        self.atol = atol # the tolerance at which we say it is converged
%%        
% \adapt{}
\textbf{\adapt{}.} The \adapt{} implementation \cite{ADAPTQSCI2023} includes several hyperparameters that control its operation. These hyperparameters have default values, which we used as our baseline for comparison.
%%
%The flag \textbf{use\_singles} determines whether single excitations are included in the operator pool; its default setting is \code{True}.
%%
\textbf{num\_pickup} (default: 100) and \textbf{coeff\_cutoff} (default: 0.001) parameters control the terms retained or removed from the Hamiltonian compressed representation.
\textbf{self\_selection} indicates whether self-selection is enabled (default: False), which forces the algorithm to work in the subspace with the correct Hamming weight. 
\textbf{iter\_max} is the maximum total number of iterations (default: 100). 
\textbf{sampling\_shots} is the number of sampling shots for measurements per iteration (default: $10^5$). 
\textbf{atol} is the absolute tolerance for convergence criteria (default: $1e-6$).
%%
%%        self.final_sampling_shots_coeff = final_sampling_shots_coeff #how many more shots to use in the calculation if the same operator appears twice or the operator parameter is close to zero
%%        self.num_precise_gradient = num_precise_gradient #how many operators from the pool to calculate the gradient more precisely 
%%        self.max_num_converged = max_num_converged #how many iterations does it need to stay within atol to be considered converged
%%        self.reset_ignored_inx_mode = reset_ignored_inx_mode #after how many iterations do we allow previously used operators to be used again
%%
\textbf{final\_sampling\_shots\_coeff} is the number of shots for the calculations if the same operator appears twice or the operator parameter is close to zero (default: 5).
\textbf{num\_precise\_gradient} is the number of operators from the pool to calculate the gradient with the full Hamiltonian, after an initial approximate calculation is carried out with the truncated Hamiltonian (default: 128).
%%
%\textbf{max\_num\_converged} specifies the maximum number of iterations within \textbf{atol} needed for a solution to be considered converged (default: 2).
%%
\textbf{reset\_ignored\_inx\_mode} specifies after how many iterations previously used operators can be reused (default: 0). 
These default values are the baseline for configuring the \adapt{} implementation, but these can be overridden (e.g.) when using \tool{} suggestions tailored per system.

%% wrapper = Wrapper(sampler,is_classical,n_qubits,n_elec,ham,ham_terms,ham_cutoff,delta_t,n_Z,alpha)
\textbf{\qcels{}. } 
%\cl{explain the hyperparameters}
The \qcels{} hyperparameters are a different set of arguments. This includes the following:
%
% \textbf{n\_elec} represents the number of electrons or particles in the system, set to be the number of qubits by default. 
%
\textbf{ham\_terms} is the number of individual terms that are retained in the Hamiltonian after truncation. The Hamiltonian is transformed to a qubit Hamiltonian formed from a linear combination of Pauli strings, the truncation then retains the largest \textbf{ham\_terms} of these and the rest are discarded. The default value is set to be 200 and we search for its optimal value between random.randint(50, 1000).
\textbf{ham\_cutoff} is the same as \textbf{coeff\_cutoff} in \adapt{} implementation, setting a minimum value for the retained coefficients, discarding all terms in the Hamiltonian with coefficients lower than this threshold.
\textbf{delta\_t} is the time step for the simulation or evolution of the system (the default is 0.03; \code{delta\_t = random.uniform(1e-3, 0.3)}), the expectation value of the time evolution operator is calculated at a set of times starting from zero and separated by this value.
\textbf{n\_Z} is the total number of points used in fitting the time evolution and so \textbf{n\_Z}$-1$ is the total number of points at which the expectation value is evaluated. The default value is 10; \code{n\_Z = random.randint(5, 25)}.
\textbf{alpha} is a scalar parameter that determines how much smaller the parameters introduced in each step of the fitting procedure are than those introduced in the previous step e.g. $r_2^{(2)}$, $\theta_2^{(2)}$ and $r_2^{(3)}$, $\theta_2^{(3)}$, are bounded by $\alpha r_1^{(1)},\pm \alpha \theta_1^{(1)}$ (default is $0.8$; \code{alpha = random.uniform(0.5, 1)}).
%%%%

\paragraph{}
Using the above, we defined different types of hyperparameter vectors per implementation.
The input vectors ('X's) were normalised to the size of vectors in 28-qubit systems and included the implementation hyperparameters and the system's Hamiltonian. The Hamiltonians were compressed by removing small elements (abs(0.05) and below). The target values (Y's) represent the predicted lowest energy levels. Note: since we utilised the classical mode when extracting data from smaller systems, these are rough approximations, not the true values.

\subsection{Experimental Setup} 
\label{sec:expsetup}
We extracted 66 files for the training phase using classical wrappers (i.e. \filesTrainAdapt{} for \adapt{} and \filesTrainAdapt{} for \qcels{}). 
%\sfo{I didn't understand what is "per implementation". If it is the 33 files, the end should move before the "for".  } % \adapt{} and \qcels{} algorithms, respectively. 
We used \textit{open-source molecular Hamiltonians}--smaller systems prepared as discussed in the data preparation paragraph in \subsectref{sec:tool}-- and \textit{the challenge Hamiltonians}--the Hamiltonians of 04 and 12 qubits for seeds \_\_00 to \_\_04 from \cite{QunaSys}.

We trained the \adapt{} with 4760 records (small dataset) and \qcels{}  with 14510 records (medium dataset), out of which 400 and 5550 records (respectively) with the challenge Hamiltonians. Per system sizes of 4-, 6-, 7-, 8-, 10-, 12-, 14-, and 16-qubit, we utilised \adaptrecA, \adaptrecB, \adaptrecC, \adaptrecD, \adaptrecE, \adaptrecF, \adaptrecG and \adaptrecH records for \adapt{} hyperparameters, and \QCELSrecA, \QCELSrecB, \QCELSrecC, \QCELSrecD, \QCELSrecE, \QCELSrecF, \QCELSrecG and \QCELSrecH for \qcels{} hyperparameters, respectively. The \qcels{} implementation ran faster than \adapt{}, allowing us to extract a medium-sized dataset for \qcels{}.
The models were trained using XGBRegressor\footnote{\href{https://xgboost.readthedocs.io/en/latest/python/python_api.html\#xgboost.XGBRegressor}{\texttt{https://xgboost.readthedocs.io/en/latest/python/python\_api.html\char"0023 xgboost.XGBRegressor}}}, version 2.1.1 of the XGBoost library,
on \gpuspec{}. We deployed the model on \specdeploy{}. 

We evaluated on a quantum simulator the predictions for several Hamiltonians of 20-, 24-, and 28-qubit systems with a known real answer (the challenge Hamiltonians were 20 and 28 qubits for seeds \_\_00 to \_\_04 \cite{QunaSys}; the rest of the seeds were open-source molecular Hamiltonians \subsectref{sec:tool}). We ran the simulations on a single virtual machine with 16 virtual CPU cores and 32 GB RAM running \code{Ubuntu 20.04.2 LTS x86\_64}. The host had a single \code{AMD EPYC 7313P CPU} (single socket, 3.0 GHz, 16 cores, 2 threads per CPU).

\subsection{Results}
\label{sec:res}

The models for \adapt{} and \qcels{} were trained with dataset sizes of \adapttrainsize{} and \qcelstrainsize{}, respectively. The model sizes were \adapttrainsizemodel{} for \adapt{} and \qcelstrainsizemodel{} for \qcels{}.

We first evaluated the model performance by calculating the mean absolute error (MAE) and the mean squared error (MSE) of two testing sets. We sampled 10\% of data from each system (first testing set) and split the remaining data into training and (second) testing sets at an 80\% - 20\% ratio. We had no validation stage due to (1) the limited number of systems with known Y's value (typically sourced from physics publications) and (2) only classically estimating the Ys for a hyperparameters assignment. Our primary focus is evaluating the algorithm \tool, not the ML component.

%% Results numbers
For \adapt{}, the MAE and MSE were \adapttenMAE and \adapttenMSE (for the 10\% set) and \adapttwentyMAE and \adapttwentyMSE (for the 20\% set), and for \qcels{}, these were \QCELStenMAE, \QCELStenMSE, \QCELStwentyMAE and \QCELStwentyMSE. These results indicate the need for a custom loss function, some feature weighting, and a validation stage with larger datasets, which we leave for future work.

% \usepackage{booktabs}
\begin{table}[ht]
\footnotesize	
\centering
\caption{Predicted optimal hyperparameters (A-J col.) with the first row stating the default values for \adapt{} implementation 
 (\textbf{A}: num\_pickup, \textbf{B}: coeff\_cutoff, \textbf{C}: self\_selection, \textbf{D}: iter\_max,	 
 \textbf{E}: sampling\_shots, \textbf{F}: atol, \textbf{G}: final\_sampling\_shots\_coeff, 	 
 \textbf{H}: num\_precise\_gradient, 
 %\textbf{I}: max\_num\_converged, 
 and \textbf{I}: reset\_ignored\_inx\_mode).}
\begin{tabular}{@{}lrrrrrrrrl@{}}
\toprule
\multicolumn{1}{c}{\textbf{System}} & \multicolumn{1}{c}{\textbf{A}} & \multicolumn{1}{c}{\textbf{B}} & \multicolumn{1}{c}{\textbf{C}} & \multicolumn{1}{c}{\textbf{D}} & \multicolumn{1}{c}{\textbf{E}} & \multicolumn{1}{c}{\textbf{F}} & \multicolumn{1}{c}{\textbf{G}} & \multicolumn{1}{c}{\textbf{H}} & \multicolumn{1}{c}{\textbf{I}} \\ \midrule
Default      & 100 & 0.001       & 0  & 100     & 100000  & 1.00E-06    & 5  & 128  & 0  \\
20qubits\_00 & 985 & 7.64647e-03 & 0  & 344658  & 49458   & 5.67168e-05 & 3  & 77   & 1  \\
20qubits\_01 & 807 & 9.41494e-03 & 1  & 382109  & 964853  & 9.00118e-05 & 6  & 79   & 64 \\
20qubits\_02 & 934 & 4.56117e-03 & 0  & 478268  & 875460  & 4.27314e-05 & 8  & 294  & 83 \\
20qubits\_03 & 551 & 1.13139e-03 & 1  & 802530  & 106374  & 7.23056e-05 & 1  & 161  & 79 \\
20qubits\_04 & 182 & 4.06265e-03 & 0  & 398085  & 61950   & 7.69414e-05 & 7  & 194  & 11 \\
20qubits\_05 & 593 & 8.46933e-04 & 1  & 278431  & 148423  & 6.66628e-05 & 7  & 158  & 58 \\
24qubits\_05 &  67 & 7.77741e-03 & 1  & 655715  & 676443  & 3.37411e-05 & 8  & 176  & 89 \\
24qubits\_06 & 292 & 4.00805e-03 & 0  & 762863  & 603105  & 6.33079e-05 & 3  & 159  & 82 \\
24qubits\_07 &  80 & 4.06535e-03 & 0  & 659716  & 711486  & 9.45728e-05 & 4  & 37   & 0  \\
24qubits\_08 & 215 & 6.59156e-03 & 1  & 885413  & 291401  & 1.74522e-05 & 1  & 247  & 81 \\
24qubits\_09 & 716 & 7.29064e-03 & 0  & 401350  & 929361  & 6.82934e-05 & 5  & 257  & 65 \\
28qubits\_00 & 197 & 1.42217e-03 & 1  & 460054  & 209518  & 7.62522e-05 & 7  & 287  & 54 \\
28qubits\_01 & 1000& 6.58567e-03 & 0  & 615787  & 300938  & 1.04770e-05 & 8  & 224  & 98 \\
28qubits\_02 & 531 & 5.88032e-03 & 1  & 402352  & 258683  & 4.81220e-06 & 8  & 277  & 47 \\
28qubits\_03 & 629 & 4.35568e-03 & 0  & 429007  & 82966   & 3.89304e-05 & 8  & 239  & 8  \\
28qubits\_04 & 698 & 7.96341e-03 & 0  & 221061  & 633412  & 7.86532e-05 & 4  & 68   & 89 \\ \bottomrule
\end{tabular}
\label{fig:tabl:hyperparams:adapt}
\end{table}

% \usepackage{booktabs}
\begin{table}[ht]
\footnotesize	
\centering
\caption{Predicted optimal hyperparameters (A-E cols.): the approximate size of the Hamiltonian as a vector of terms, 
%% delta_t, n_Z, ham_terms, ham_cutoff, alpha]
\textbf{A}: delta\_t,
\textbf{B}: n\_Z,
\textbf{C}: ham\_terms, \textbf{D}: ham\_cutoff, 
and \textbf{E}: alpha).}

\begin{tabular}{@{}lrrrrr@{}}
\toprule
\multicolumn{1}{c}{\textbf{System}} & \multicolumn{1}{c}{\textbf{A}} & \multicolumn{1}{c}{\textbf{B}} & \multicolumn{1}{c}{\textbf{C}} & \multicolumn{1}{c}{\textbf{D}} & \multicolumn{1}{c}{\textbf{E}}  \\ \midrule
Default      & 0.03         & 10 & 200 & 1e-9         & 0.8   \\
20qubits\_00 & 1.37929e-01  & 14 & 877 & 6.77275e-03  & 9.76144e-01 \\
20qubits\_01 & 2.87833e-01  & 8  & 69  & 9.23310e-04  & 5.26429e-01 \\
20qubits\_02 & 1.05366e-01  & 10 & 989 & 4.31098e-04  & 8.77018e-01 \\
20qubits\_03 & 1.04235e-01  & 16 & 68  & 6.91101e-03  & 5.12876e-01 \\
20qubits\_04 & 1.31645e-01  & 19 & 971 & 2.10235e-03  & 7.76523e-01 \\
20qubits\_05 & 1.98379e-01  & 7  & 187 & 8.34526e-03  & 5.33536e-01 \\
24qubits\_05 & 2.08699e-01  & 24 & 933 & 5.75035e-03  & 6.20185e-01 \\
24qubits\_06 & 1.75486e-01  & 10 & 398 & 7.28888e-03  & 9.72887e-01 \\
24qubits\_07 & 1.80166e-01  & 11 & 286 & 1.06348e-03  & 9.75924e-01 \\
24qubits\_08 & 2.72722e-01  & 22 & 453 & 6.67156e-03  & 8.29813e-01 \\
24qubits\_09 & 2.09069e-01  & 24 & 650 & 3.99377e-03  & 5.20032e-01 \\
28qubits\_00 & 7.00174e-03  & 7  & 310 & 6.71154e-03  & 6.32763e-01 \\
28qubits\_01 & 1.99713e-01  & 23 & 309 & 8.39957e-03  & 5.54053e-01 \\
28qubits\_02 & 5.46760e-02  & 12 & 997 & 1.22706e-03  & 5.60243e-01 \\
28qubits\_03 & 4.23353e-02  & 6  & 553 & 8.40543e-03  & 8.66162e-01 \\
28qubits\_04 & 2.33005e-01  & 25 & 280 & 1.30071e-04  & 5.98526e-01 \\ 
\bottomrule
\end{tabular}
\label{fig:tabl:hyperparams:qcels}
\end{table}

\begin{table}[ht]
\centering
\footnotesize	
\caption{
Hamiltonian sizes of seven Hamiltonian representations.
The Hamiltonian sizes in the five rightmost columns are post-Jordan-Wigner Transformation.
We marked the open-source molecular Hamiltonians with an asterisk.
(1) FermionOperator term count, in the original Hamiltonian representation (\#terms col.);
(2) The size of a vector of floating point numbers in NumPy of the compressed, flattened FermionOperator representation for machine learning processing (\#floats in ML model col.);
(3) initial Hamiltonian term count (Init. col.), and its size compressed via cutoff and term selection of (4) default and (5) optimised \adapt{}, and (6) default and (7) optimised QCELS hyperparameters setup (Round ADAPT-QSCI, Round opt. ADAPT-QSCI, Truncated QCELS, and Truncated opt. QCELS columns).
}
\label{tbl:size:vs:comp}

\begin{tabular}{lrr||rrrrr}
\cline{4-8}
& \multicolumn{1}{l}{}            & \multicolumn{1}{l|}{}            & \multicolumn{5}{c|}{\cellcolor[HTML]{C0C0C0}\textbf{\#terms in Hamiltonian (Jordan-Wigner)}} \\ \hline
\rowcolor[HTML]{EFEFEF} 
\multicolumn{1}{c}{\cellcolor[HTML]{EFEFEF}\textbf{System}} & \multicolumn{1}{c}{\cellcolor[HTML]{EFEFEF}\textbf{\#terms}} & \multicolumn{1}{c}{\cellcolor[HTML]{EFEFEF}\textbf{\begin{tabular}[c]{@{}c@{}}\#floats in \\ ML model\end{tabular}}} & \multicolumn{1}{c}{\cellcolor[HTML]{EFEFEF}Init.} & \multicolumn{1}{c}{\cellcolor[HTML]{EFEFEF}\begin{tabular}[c]{@{}c@{}}Round\\ ADAPT-QSCI\end{tabular}} & \multicolumn{1}{c}{\cellcolor[HTML]{EFEFEF}\begin{tabular}[c]{@{}c@{}}Round, Opt.,\\ ADAPT-QSCI\end{tabular}} & \multicolumn{1}{c}{\cellcolor[HTML]{EFEFEF}\begin{tabular}[c]{@{}c@{}}Truncated\\ QCELS\end{tabular}} & \multicolumn{1}{c}{\cellcolor[HTML]{EFEFEF}\begin{tabular}[c]{@{}c@{}}Truncated, \\ Opt.,QCELS\end{tabular}} \\ \hline

20qubits\_00    & 63636 & 91  & 14537 & 100 & 985  & 200 & 877   \\
20qubits\_01    & 54717 & 287 & 14537 & 100 & 807  & 200 & 69    \\
20qubits\_02    & 54723 & 445 & 14535 & 100 & 934  & 200 & 989   \\
20qubits\_03    & 54710 & 338 & 14551 & 100 & 551  & 200 & 68    \\
20qubits\_04    & 63636 & 65  & 14537 & 100 & 182  & 200 & 971   \\
20qubits\_05(*) & 46    & 46  & 67    & 67  & 67   & 66  & 66    \\
24qubits\_05(*) & 56    & 48  & 81    & 81  & 67   & 80  & 80    \\
24qubits\_06(*) & 56    & 46  & 81    & 81  & 81   & 80  & 80    \\
24qubits\_07(*) & 56    & 50  & 81    & 81  & 80   & 80  & 80    \\
24qubits\_08(*) & 56    & 51  & 81    & 81  & 81   & 80  & 80    \\
24qubits\_09(*) & 56    & 47  & 81    & 81  & 81   & 80  & 80    \\
28qubits\_00    & 98700 & 167 & 42982 & 100 & 197  & 200 & 310   \\
28qubits\_01    & 98700 & 158 & 42982 & 100 & 1000 & 200 & 309   \\
28qubits\_02    & 98700 & 98  & 42982 & 100 & 531  & 200 & 997   \\
28qubits\_03    & 98700 & 144 & 42982 & 100 & 629  & 200 & 553   \\
28qubits\_04    & 98700 & 213 & 42982 & 100 & 698  & 200 & 280   \\ \hline

\end{tabular}
\end{table}

%% (Size: the approximate size of the Hamiltonian as a vector after compression
% System | Size | Compressed
% 20qubits_00 | 63636 | 91
% 20qubits_01 | 54717 | 287
% 20qubits_02 | 54723 | 445
% 20qubits_03 | 54710 | 338
% 20qubits_04 | 63636 | 65
% 28qubits_00 | 98700 | 167
% 28qubits_01 | 98700 | 158
% 28qubits_02 | 98700 | 98
% 28qubits_03 | 98700 | 144
% 28qubits_04 | 98700 | 213

% System | Size | Compressed
% 20qubits_05 | 46 | 46
% 24qubits_05 | 56 | 48
% 24qubits_06 | 56 | 46
% 24qubits_07 | 56 | 50
% 24qubits_08 | 56 | 51
% 24qubits_09 | 56 | 47
\paragraph{Optimal Hyperparameters Prediction Results. } 

\autoref{fig:tabl:hyperparams:adapt} and \autoref{fig:tabl:hyperparams:qcels} show the optimal hyperparameters found by \tool with each Hamiltonian (System col.) employing the two models above. The first row shows the default values of the implementation. 
%% Descibe each Table:
\autoref{fig:tabl:hyperparams:adapt} shows the \adapt{}'s optimal hyperparameters, with col. A-I represents the predicted optimal values of the hyperparameters. Similarly, \autoref{fig:tabl:hyperparams:qcels} presents the optimal values for the hyperparameters of the \qcels{} implementation in col. A-E.
%% maximum number of shots: 10000000
%We then normalised the 
%iter\_max to have the same limit as in the default parameters (that is, iter\_max=1e7/sampling\_shots) because predicted iter\_max was a six-figure number, which partly defeats the purpose of using machine learning for optimization if we have unlimited resources. 
The value presented in \autoref{fig:tabl:hyperparams:adapt} for the predicted iter\_max was capped during execution\footnote{The platform automatically stopped the computation once the maximum number of shots, $10\,000\,000$, was reached.} 
to ensure that we do not have unlimited resources. In practice, the iter\_max has the same limit as with the default parameters (i.e. iter\_max=1e7/sampling\_shots using the values in col. E and col. D). 

The results in \autoref{fig:tabl:hyperparams:adapt} suggest that the optimisation opted to (sometimes) lower the performance of each iteration, resulting in less precise results per iteration but using more iterations overall by reducing (sometimes) sampling shots and increasing the atol, coeff\_cutoff and iter\_max. However, when the optimisation predicted both iter\_max and sampling\_shots at maximum values, ignoring their correlation, the execution was capped, resulting in fewer iterations in practice.
The results in \autoref{fig:tabl:hyperparams:qcels} show that the optimisation favoured increasing the number of large coefficients (except for 20qubits\_01 and 20qubits\_03) using larger ham\_cutoff values and, hence, leaving fewer small coefficients. Further, the optimisation selected higher values for n\_z and delta\_t, with no clear preference for alpha.
%\todo{Connor: do you have something to add about alpha and delta? Nz is commonly higher than 10.}

The hyperparameters affect the number of terms considered during the QE implementation executions, leading to different cutoffs, the number of terms to select and compression rates used for the ML model and the QE implementation.
\autoref{tbl:size:vs:comp} presents an analysis of these sizes under different choices of values of the hyperparameters for the ML model and both QE implementations. 
The Hamiltonian sizes in the five rightmost columns in \autoref{tbl:size:vs:comp} are post-Jordan-Wigner Transformation.
Per Hamiltonian (System col.), we recorded the FermionOperator term count in the original Hamiltonian representation (\#terms col.), the initial Hamiltonian term count in Jordan Wigner form (Init. col.), and this representation's size when compressed via cutoff and term selection of the default and optimised \adapt{}, and the default and optimised QCELS hyperparameters setup (Round ADAPT-QSCI col., Round, opt., ADAPT-QSCI col., Truncated QCELS col. and Truncated, opt., QCELS col., respectively).
The column \textit{\#floats in ML model} is the size of a vector of floating point numbers in NumPy of the compressed, flattened FermionOperator representation for machine learning processing. We construct this vector as follows.
The Hamiltonian, initially expressed as a collection of terms with associated coefficients, is transformed into a flat numerical vector to streamline processing and analysis. This transformation involves extracting the coefficients, filtering out those with negligible contributions, and selecting key elements significant enough to influence the overall Hamiltonian. The resulting coefficients are then organised into a consistent and compact format. This representation retains the essential numerical properties of the Hamiltonian while ensuring uniformity and suitability for downstream processing.
Lastly, we marked the open-source molecular Hamiltonians with an asterisk in \autoref{tbl:size:vs:comp}. The source of the Hamiltonians is described in \autoref{sec:expsetup}.

In \autoref{tbl:size:vs:comp}, the number of terms in the open-source molecular Hamiltonian systems ($\leq 46$) is smaller by several orders of magnitude compared to the challenge Hamiltonians ($\geq 54k$).
Even after compression, the open-source molecular Hamiltonians systems were still much smaller on average. This suggests that these systems may differ to some extent. 

The next section examines the differences in the percentage of error achieved by each system, both with and without optimised hyperparameters, and in relation to their sizes.

%%%%%%%% LARGER SYSTEM RUNS

\paragraph{Results of Execution with Different Hyperparameters. } \autoref{tbl:main:res1} and \autoref{tbl:main:res2} summarise the results of executing \adapt{} and \qcels{} on a quantum simulator with the default and optimal hyperparameters in \autoref{fig:tabl:hyperparams:adapt} and \autoref{fig:tabl:hyperparams:qcels}. 
%%%
We evaluated predictions on several Hamiltonians of 20-, 24- and 28-qubit systems (System col.); we excluded the results of \qcels{} on 28 qubits from \autoref{tbl:main:res1} and discuss these in detail in the discussion paragraph below.
%%
%%  with a known solution (Value col.)
\autoref{tbl:main:res1} presents the result obtained with the challenge Hamiltonians, while \autoref{tbl:main:res2} is the results using the open-source molecular Hamiltonians. 
Both tables present, for each algorithm and set of hyperparameters (Algorithm col.), the task score (Task Score col.), 
%the approximate run time (Approx. Runtime col.), 
and the number of iterations completed (Itr. col.).  
For \autoref{tbl:main:res1}, the true results (Value col.) are known \cite{QAGC2024}, and we computed the relative error (Error (\%) col.) per experiment.
%% Original text:
% Both tables present, for each algorithm and set of hyperparameters (Algorithm col.), the relative error (Error (\%) col.), the task score (Task Score col.), the approximate run time (Approx. Runtime col.), and the number of iterations completed (Itr. col.). We ran each experiment several times and selected the best result.
Note that in total, we evaluated our prototype framework using 16 systems, each tested with two implementations (\adapt and \qcels) under default and optimised hyperparameters, resulting in 16×2×2=64 experiments. Each experiment was repeated 10 times, and the best (lowest) result from these repetitions was selected\footnote{For \qcels{}, the implementation frequently crashed in the machine during our experiments, likely due to insufficient resources, making it challenging to obtain results. To mitigate this, we adopted a resilient execution strategy, running additional repetitions and collecting results only from the first 10 non-crashing executions. In some cases, we were fortunate, and all executions completed without crashing; in these cases, we excluded the last machine’s results to maintain consistency across experiments.}.
While the time per repeat is excluded from this paper and tables due to unreliable recording on our host machine, timing logs are available in our Zenodo record \cite{bensoussan_2024_artifact_ML2QarXiv}.

\begin{table}[t!]
\centering
\scriptsize
\caption{The challenge Hamiltonians: Results of execution of \adapt{} and \qcels{} with default and optimal hyperparameters (Itr. is Iterations Completed when the result was obtained.).}
\label{tbl:main:res1}
%%%%%%%%%%%%%%%%%%%%%%%%%%%%%%%%%%%%%%%%%%%%%%%%%%%%%%%%%%%%%%%%%%%%%%%%%%%
%%%%%%%%%%%%%%%%%%%%%%%%%%%%%%%%%%%%%%%%%%%%%%%%%%%%%%%%%%%%%%%%%%%%%%%%%%%
%%%%%%%%%%%%%%%%%%%%%%%%%%%%%%%%%%%%%%%%%%%%%%%%%%%%%%%%%%%%%%%%%%%%%%%%%%%
%%%%%%%%%%%%%%%%%%%%%%%%%%%%%%%%%%%%%%%%%%%%%%%%%%%%%%%%%%%%%%%%%%%%%%%%%%%
%%%%%%%%%%%%%%%%%%%%%%%%%%%%%%%%%%%%%%%%%%%%%%%%%%%%%%%%%%%%%%%%%%%%%%%%%%%
%%%%%%%%%%%%%%%%%%%%%%%%%%%%%%%%%%%%%%%%%%%%%%%%%%%%%%%%%%%%%%%%%%%%%%%%%%%
%%%%%%%%%%%%%%%%%%%%%%%%%%%%%%%%%%%%%%%%%%%%%%%%%%%%%%%%%%%%%%%%%%%%%%%%%%%
%%%%%%%%%%%%%%%%%%%%%%%%%%%%%%%%%%%%%%%%%%%%%%%%%%%%%%%%%%%%%%%%%%%%%%%%%%%
%%%%%%%%%%%%%%%%%%%%%%%%%%%%%%%%%%%%%%%%%%%%%%%%%%%%%%%%%%%%%%%%%%%%%%%%%%%
%%%%%%%%%%%%%%%%%%%%%%%%%%%%%%%%%%%%%%%%%%%%%%%%%%%%%%%%%%%%%%%%%%%%%%%%%%%
\begin{tabular}{@{}lllrlr@{}}
\toprule %% 100*(-20.294244005205 + 22.046059902) / 22.046059902
% (2.42+6.27+5.79+3.24+3.30)/5 - ADAPT opt 20 4.204
% (4.28+4.65+4.28+4.49+4.59)/5 - ADAPT default 20 4.458

% (8.17+0.83+8.43+6.79+8.53)/5 - QCELS opt 20
% (10.2+7.64+8.47+8.81+7.05)/5 - QCELS default 20 8.434

% (6.33+6.37+6.38+6.11+6.78)/5 - ADAPT opt 28 6.39
% (6.44+6.36+6.60+6.44+6.71)/5 - ADAPT default 28 6.51
\textbf{System}             & \textbf{Value}                   & \textbf{Algorithm}  & \textbf{Error (\%)} & \textbf{Task Score}              & \textbf{Itr.}        \\ \midrule
\multirow{4}{*}{20qubits\_00} & \multirow{4}{*}{-22.046059902} & ADAPT-QSCI, Opt.    & 2.42                & \textbf{-21.5134481079129}       & 147                  \\
                            &                                  & ADAPT-QSCI, Default & 4.28                & -21.102120951277936              & 85                   \\
                            &                                  & QCELS, Opt.         & 8.17                & -20.244706524965135              & 13                   \\
                            &                                  & QCELS, Default      & 10.2                & -19.79974257492709               & 9                    \\ \midrule
%%%%%%%
\multirow{4}{*}{20qubits\_01} & \multirow{4}{*}{-22.046059902} & ADAPT-QSCI, Opt.    & 6.27                & -20.66347471739218               & 10                   \\
                            &                                  & ADAPT-QSCI, Default & 4.65                & -21.02122671542096               & 78                   \\
                            &                                  & QCELS, Opt.         & 0.83                & \textbf{-21.86260739785393}      & 7                    \\
                            &                                  & QCELS, Default      & 7.64                & -20.361475620508653              & 9                    \\ \midrule
%%%%%%%%
\multirow{4}{*}{20qubits\_02} & \multirow{4}{*}{-22.046059902} & ADAPT-QSCI, Opt.    & 5.79                & -20.76850360872923               & 11                   \\
                            &                                  & ADAPT-QSCI, Default & 4.28                & \textbf{-21.101500354754492}     & 65                   \\
                            &                                  & QCELS, Opt.         & 8.43                & -20.18686453079191               & 9                   \\
                            &                                  & QCELS, Default      & 8.47                & -20.178731347808355              & 9                    \\ \midrule
%%%%%%%%
\multirow{4}{*}{20qubits\_03} & \multirow{4}{*}{-22.046059902} & ADAPT-QSCI, Opt.    & 3.24                & \textbf{-21.33253115024456}      & 93                   \\
                            &                                  & ADAPT-QSCI, Default & 4.49                & -21.055181221011736              & 79                   \\
                            &                                  & QCELS, Opt.         & 6.79                & -20.548184887148853              & 15                   \\
                            &                                  & QCELS, Default      & 8.81                & -20.104190198375786              & 9                    \\ \midrule
%%%%%%%%
\multirow{4}{*}{20qubits\_04} & \multirow{4}{*}{-22.046059902} & ADAPT-QSCI, Opt.    & 3.30                & \textbf{-21.31780361442556}      & 126                  \\
                            &                                  & ADAPT-QSCI, Default & 4.59                & -21.034981116395308              & 83                   \\
                            &                                  & QCELS, Opt.         & 8.53                & -20.16538489337923               & 15                  \\
                            &                                  & QCELS, Default      & 7.05                & -20.49223412640405               & 9                   \\ \midrule
%%%%%%%%
\multirow{2}{*}{28qubits\_00} & \multirow{2}{*}{-30.748822808} & ADAPT-QSCI, Opt.    & 6.33                & \textbf{-28.80238180353666}      & 45                   \\
                            &                                  & ADAPT-QSCI, Default & 6.44                & -28.76881293227412               & 58                   \\
                            %&                                  & QCELS, Opt.      %   & 6.35                 & -28.794794630409676       %       & 48                    \\
                            %&                                  & QCELS, Default   %   & 6.40                 & -28.779806038153325       %       & 65                   \\ 
                            \midrule
%%%%%%%%
\multirow{2}{*}{28qubits\_01} & \multirow{2}{*}{-30.748822808} & ADAPT-QSCI, Opt.    & 6.37                & -28.788828203287586              & 33                   \\
                            &                                  & ADAPT-QSCI, Default & 6.36                & \textbf{-28.79342085633283}      & 76                   \\
                            %&                                  & QCELS, Opt.      %   & 6.41                 & -28.777776815200454       %       & 34                   \\
                            %&                                  & QCELS, Default   %   & 6.46                 & -28.763094962187317       %       & 74                   \\ 
                            \midrule
%%%%%%%%
\multirow{2}{*}{28qubits\_02} & \multirow{2}{*}{-30.748822808} & ADAPT-QSCI, Opt.    & 6.38                & \textbf{-28.788448328499076}     & 38                   \\
                            &                                  & ADAPT-QSCI, Default & 6.60                & -28.720685906958145              & 59                   \\
                            %&                                  & QCELS, Opt.      %   & 6.39                 & -28.7835710030780         %       & 39                   \\
                            %&                                  & QCELS, Default   %   & 6.66                 & -28.70067301821086        %       & 57                   \\ 
                            \midrule
%%%%%%%%
\multirow{2}{*}{28qubits\_03} & \multirow{2}{*}{-30.748822808} & ADAPT-QSCI, Opt.    & 6.11                & \textbf{-28.871013405346638}     & 56                   \\
                            &                                  & ADAPT-QSCI, Default & 6.44                & -28.767725361229388              & 75                   \\
                            %&                                  & QCELS, Opt.      %   & 6.18                 & -28.848400001238723       %       & 53                   \\
                            %&                                  & QCELS, Default   %   & 6.66                 & -28.699830076240456              & 70                   \\ 
                            \midrule
%%%%%%%%
\multirow{2}{*}{28qubits\_04} & \multirow{2}{*}{-30.748822808} & ADAPT-QSCI, Opt.    & 6.78                & -28.66362404354463               & 15                   \\
                            &                                  & ADAPT-QSCI, Default & 6.71                & \textbf{-28.68471298123107}               & 37                   \\
                            %&                                  & QCELS, Opt.        % & 6.78                 & -28.66351868143365               & 16                   \\
                            %&                                  & QCELS, Default      & 6.70                 & \textbf{-28.689855023621764}     & 45                   \\ 
                            \bottomrule
                            %% 100*(-28.689855023621764 + 30.748822808) / -30.748822808
\end{tabular}         
\end{table}
%%%%%%%%%%%%%%%%%%%%%%%%%%%%%%%%%%%%%%%%%%%%%%%%%%%%%%%%%%%%%%%%%%%%%%%%%%%
%%%%%%%%%%%%%%%%%%%%%%%%%%%%%%%%%%%%%%%%%%%%%%%%%%%%%%%%%%%%%%%%%%%%%%%%%%%
%%%%%%%%%%%%%%%%%%%%%%%%%%%%%%%%%%%%%%%%%%%%%%%%%%%%%%%%%%%%%%%%%%%%%%%%%%%
%%%%%%%%%%%%%%%%%%%%%%%%%%%%%%%%%%%%%%%%%%%%%%%%%%%%%%%%%%%%%%%%%%%%%%%%%%%
%%%%%%%%%%%%%%%%%%%%%%%%%%%%%%%%%%%%%%%%%%%%%%%%%%%%%%%%%%%%%%%%%%%%%%%%%%%
%%%%%%%%%%%%%%%%%%%%%%%%%%%%%%%%%%%%%%%%%%%%%%%%%%%%%%%%%%%%%%%%%%%%%%%%%%%
%%%%%%%%%%%%%%%%%%%%%%%%%%%%%%%%%%%%%%%%%%%%%%%%%%%%%%%%%%%%%%%%%%%%%%%%%%%
%%%%%%%%%%%%%%%%%%%%%%%%%%%%%%%%%%%%%%%%%%%%%%%%%%%%%%%%%%%%%%%%%%%%%%%%%%%
%%%%%%%%%%%%%%%%%%%%%%%%%%%%%%%%%%%%%%%%%%%%%%%%%%%%%%%%%%%%%%%%%%%%%%%%%%%
%%%%%%%%%%%%%%%%%%%%%%%%%%%%%%%%%%%%%%%%%%%%%%%%%%%%%%%%%%%%%%%%%%%%%%%%%%%
%%%%%%%%%%%%%%%%%%%%%%%%%%%%%%%%%%%%%%%%%%%%%%%%%%%%%%%%%%%%%%%%%%%%%%%%%%%
%%%%%%%%%%%%%%%%%%%%%%%%%%%%%%%%%%%%%%%%%%%%%%%%%%%%%%%%%%%%%%%%%%%%%%%%%%%
%%%%%%%%%%%%%%%%%%%%%%%%%%%%%%%%%%%%%%%%%%%%%%%%%%%%%%%%%%%%%%%%%%%%%%%%%%%
%%%%%%%%%%%%%%%%%%%%%%%%%%%%%%%%%%%%%%%%%%%%%%%%%%%%%%%%%%%%%%%%%%%%%%%%%%%
%%%%%%%%%%%%%%%%%%%%%%%%%%%%%%%%%%%%%%%%%%%%%%%%%%%%%%%%%%%%%%%%%%%%%%%%%%%
%%%%%%%%%%%%%%%%%%%%%%%%%%%%%%%%%%%%%%%%%%%%%%%%%%%%%%%%%%%%%%%%%%%%%%%%%%%

\begin{table}[t!]
\centering
\scriptsize
\caption{Open-source molecular Hamiltonians: Results of execution of \adapt{} and \qcels{} with default and optimal hyperparameters (Itr. is Iterations Completed). We are not providing the exact ground state energy values for reference due to the significant computational difficulty in obtaining them for these systems. Instead, we present the energy values obtained from \adapt{}, which serve as a reliable lower bound for the ground state energy.
For \qcels{}, we fixed the task score by a constant factor that was missing in the original computation.}
\label{tbl:main:res2}
%%%%%%%%%%%%%%%%%%%%%%%%%%%%%%%%%%%%%%%%%%%%%%%%%%%%%%%%%%%%%%%%%%%%%%%%%%%
%%%%%%%%%%%%%%%%%%%%%%%%%%%%%%%%%%%%%%%%%%%%%%%%%%%%%%%%%%%%%%%%%%%%%%%%%%%
%%%%%%%%%%%%%%%%%%%%%%%%%%%%%%%%%%%%%%%%%%%%%%%%%%%%%%%%%%%%%%%%%%%%%%%%%%%
%%%%%%%%%%%%%%%%%%%%%%%%%%%%%%%%%%%%%%%%%%%%%%%%%%%%%%%%%%%%%%%%%%%%%%%%%%%
%%%%%%%%%%%%%%%%%%%%%%%%%%%%%%%%%%%%%%%%%%%%%%%%%%%%%%%%%%%%%%%%%%%%%%%%%%%
%%%%%%%%%%%%%%%%%%%%%%%%%%%%%%%%%%%%%%%%%%%%%%%%%%%%%%%%%%%%%%%%%%%%%%%%%%%
%%%%%%%%%%%%%%%%%%%%%%%%%%%%%%%%%%%%%%%%%%%%%%%%%%%%%%%%%%%%%%%%%%%%%%%%%%%
%%%%%%%%%%%%%%%%%%%%%%%%%%%%%%%%%%%%%%%%%%%%%%%%%%%%%%%%%%%%%%%%%%%%%%%%%%%
%%%%%%%%%%%%%%%%%%%%%%%%%%%%%%%%%%%%%%%%%%%%%%%%%%%%%%%%%%%%%%%%%%%%%%%%%%%
%%%%%%%%%%%%%%%%%%%%%%%%%%%%%%%%%%%%%%%%%%%%%%%%%%%%%%%%%%%%%%%%%%%%%%%%%%%
\begin{tabular}{@{}llll@{}}
\toprule
 %% 100*(-8.622105677825234   + 22.046059902) / -22.046059902
\textbf{System}              & \textbf{Algorithm}   & \textbf{Task Score}             & \textbf{Itr.} \\ \midrule
\multirow{4}{*}{20qubits05}  & ADAPT-QSCI, Opt.      & -13.308108432842927            & 26  \\
                             & ADAPT-QSCI, Default   & \textbf{-14.000000000000021}   & 9   \\
                             & QCELS, Opt.           & -13.602690623110092            & 6   \\
                             & QCELS, Default        & -13.57495914514915             & 9   \\ \midrule
%%%%%%%%
\multirow{4}{*}{24qubits05}  & ADAPT-QSCI, Opt.      & -8.965984985383866             & 10  \\
                             & ADAPT-QSCI, Default   & \textbf{-9.667517470048752}    & 40  \\
                             & QCELS, Opt.           & -7.987619347414383             & 23  \\
                             & QCELS, Default        & -8.23727545736091              & 9  \\ \midrule
%%%%%%%%
\multirow{4}{*}{24qubits06}  & ADAPT-QSCI, Opt.      & \textbf{-9.460283005059704}    & 7  \\
                             & ADAPT-QSCI, Default   & -9.460283005059702             & 5   \\
                             & QCELS, Opt.           & -9.35586373343566              & 9   \\
                             & QCELS, Default        & -9.342675357790028             & 9   \\ \midrule
%%%%%%%%
\multirow{4}{*}{24qubits07}  & ADAPT-QSCI, Opt.     & -7.570508389218553              & 12  \\
                             & ADAPT-QSCI, Default  & \textbf{-9.965785682971944}     & 27  \\
                             & QCELS, Opt.           & -7.7996489422244103            & 10  \\
                             & QCELS, Default        & -7.6227840132533573            & 9  \\ \midrule
%%%%%%%%
\multirow{4}{*}{24qubits08}  & ADAPT-QSCI, Opt.     & -6.277553305948219              & 27  \\
                             & ADAPT-QSCI, Default  & \textbf{-7.74244754268563}      & 30  \\
                             & QCELS, Opt.           & -5.1411838474119161            & 21  \\
                             & QCELS, Default        & -5.2184341011492506            & 9  \\ \midrule
%%%%%%%%
\multirow{4}{*}{24qubits09}  & ADAPT-QSCI, Opt.     & -13.136457290516129             & 5   \\
                             & ADAPT-QSCI, Default  & \textbf{-13.390021054767033}    & 14   \\
                             & QCELS, Opt.           & -13.13645729051612             & 23   \\
                             & QCELS, Default        & -12.799039886601325            & 9   \\ \bottomrule
%%%%%%%%
\end{tabular}
\end{table}

%% Present the results for Table 4
In \autoref{tbl:main:res1}, the optimised hyperparameters tended to achieve better results than the default. The optimised \adapt{} found the best solution for 6 systems, optimised \qcels{} for 1, default \adapt{} for 3, and default \qcels{} for none, out of 10 tested systems. 
%%%
For 20-qubit systems, optimised \adapt{} had the lowest error at 4.2\%, followed by default \adapt{} at 4.46\%.
Optimised and default \qcels{} had average error rates of 6.55\% and 8.43\%, respectively. 
For 28-qubit systems, optimised \adapt{} had the lowest error at 6.39\%, followed by default \adapt{} at 6.51\%.
%% Present the results for Table 5:
\autoref{tbl:main:res2} shows that default parameters performed better. The default \adapt{} found the best solution for 5 systems, followed by optimised \adapt{} for 1, out of 6 tested systems. %\todo{Elena: add error rate.} 

% (2.42+6.27+5.79+3.24+3.30)/5 - ADAPT opt 20 4.204
% (4.28+4.65+4.28+4.49+4.59)/5 - ADAPT default 20 4.46

% (8.17+0.83+8.43+6.79+9)/5 - QCELS opt 20
% (10.2+7.64+8.47+8.81+7.05)/5 - QCELS default 20 8.43

% (6.33+6.37+6.38+6.11+6.78)/5 - ADAPT opt 28 6.39
% (6.44+6.36+6.60+6.44+6.71)/5 - ADAPT default 28 6.51

%% \textbf{RQ1:} To what extent can \tool's optimisation of hyperparameters alone accelerate and improve the efficiency and accuracy of Quantum Eigensolver (QE) algorithms on NISQ devices in terms of runtime, error, and system size?
\begin{tcolorbox}[colback=gray!20, colframe=black]
    \textbf{RQ1 Answer. }
The results indicate a limited ability to improve solely through hyperparameter optimisation, though the optimised hyperparameters performed better in \autoref{tbl:main:res1}. A more refined model based on Hamiltonian characteristics is required to assess the optimisation of hyperparameters' impact on accelerating and improving QE efficiency and accuracy.
\end{tcolorbox}

RQ1 answer in detail:  \textbf{(20-qubit)} Optimised setups showed larger error variability (0.83\%–8.53\%) than default setups (4.28\%–10.2\%). While the \qcels{} optimised version achieved the lowest error (0.83\%) with just seven iterations, the \adapt{} optimised version performed better in other cases. Both implementations, with default vs optimised hyperparameters, had similar iteration counts, with \qcels{} requiring fewer iterations in general. However, optimised setups mostly performed worse when the number of iterations was too low. 
%%%%
\textbf{(24-qubit)} \adapt{} default outperformed the rest of the configurations, with no particular further observations. It suggests that optimisation did not enhance accuracy for systems whose Hamiltonians have a relatively small number of terms.
%%%%
\textbf{(28-qubit)} The number of iterations was almost halved on average, with optimised setups requiring 37 compared to 61 iterations with default setups, suggesting potential acceleration benefits. 
Furthermore, optimised \adapt{}, when outperforming the default setup, achieved lower errors—often by an order of magnitude—while default setups only had marginal advantages (0.x vs 0.0x error differences). This suggests that hyperparameter optimisation can yield substantial accuracy gains despite not always outperforming default settings.
%%%
\textbf{For all systems}, \adapt{} optimised hyperparameters used 41 iterations versus 51 for the default setup, with the opposite trend for \qcels, whose runs with optimised hyperparameters used 14 iterations versus 9 with the default setup. Eight systems performed better with optimised setups, mostly from \autoref{tbl:main:res1}, while eight performed better with the default setups, mostly from \autoref{tbl:main:res2}.

%% This is for RQ2:
%% 
%% \textbf{RQ2:} How scalable are machine learning-predicted hyperparameters learned on smaller systems when applied to QE algorithms for Hamiltonian systems with increasing qubit number and complexity?
%% Discussing diff in prediction ability

\begin{tcolorbox}[colback=gray!20, colframe=black]
    \textbf{RQ2 Answer. }
 Our understanding of \tool's scalability remains somewhat limited: 70\% of the challenge Hamiltonians' scores improved with optimised setup along with a general decrease of the error in general but showed little improvement in open-source molecular Hamiltonians. This suggests that other factors, like Hamiltonian nature, hyperparameter types, padding, or even data proportions (per source of Hamiltonians), affect scalability and require further investigation.
\end{tcolorbox}

RQ2 answer in detail: We considered here two factors: (1) training data size and (2) prior hyperparameter optimisation.
The challenge Hamiltonian systems contributed 8.4\% of \adapt{}'s and 37.9\% of \qcels{}'s training data, with \qcels{} using much shorter vectors. Open-source molecular Hamiltonians averaged 48 elements, compared to 200.6 for the challenge Hamiltonian (\autoref{tbl:size:vs:comp}, \textit{\#floats in ML model} col.).
\autoref{tbl:main:res1} shows that optimised hyperparameters performed better for the challenge Hamiltonians. We observed the opposite for the open-source molecular Hamiltonians in \autoref{tbl:main:res2}. 
This suggests that training data quantity alone is not the primary factor but rather a feature dimension—as open-source molecular Hamiltonians had fewer terms in the ML phase, which may affect the model’s ability to generalise.
Finally, the default \qcels{} performed the worst, as its original hyperparameters had not been previously optimised by other methods, possibly explaining its weaker performance in comparison to the challenge algorithm, \adapt{}.

\begin{figure}[ht]
    \centering
    \begin{subfigure}[b]{0.45\linewidth}
        \centering
        \includegraphics[width=\linewidth]{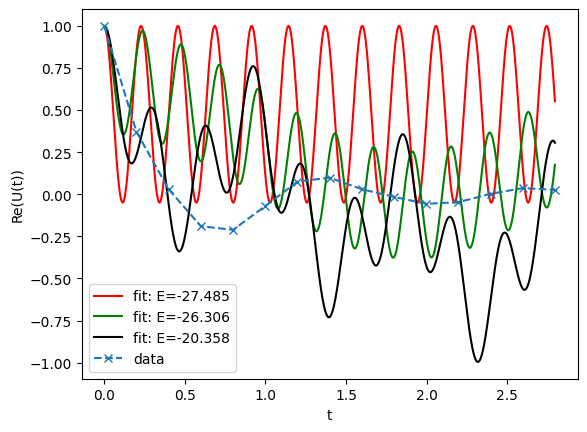}
        \caption{The original high-frequency fitting procedure with a period of approximately the spacing between the data. The damping of the oscillations is observed in the data collected from the simulated quantum algorithm.}
        \label{fig:28qbs_faulty}
    \end{subfigure}
    \hfill
    \begin{subfigure}[b]{0.45\linewidth}
        \centering
        \includegraphics[width=\linewidth]{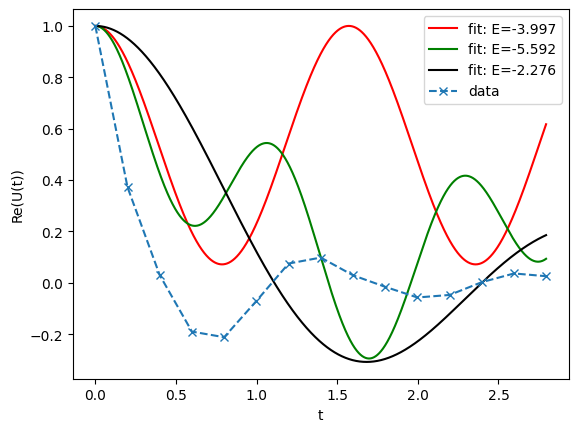}
        \caption{An updated fitting procedure which prevents high frequency fits, but does not significantly improve the results in this case because of the error in the collected time evolution data.}
        \label{fig:28qbs_sort_of_fixed}
    \end{subfigure}
    \hfill
\end{figure}

\paragraph{Discussion.} \color{black} Analysis of the scalability of the optimisation procedure is hampered by the limitations of classical simulators. For larger system sizes, the time to evaluate a large circuit can quickly become prohibitive for running many iterations of algorithms, even with access to large compute resources and sufficient memory to store the system's state. These challenges prevented the successful evaluation of the \qcels{} algorithm at $28$ qubits with a strong damping effect on the oscillations of time evolving expectation value. Possible causes of such an effect could include the Trotter error \cite{trotter,Suzuki1976} introduced by the implementation of the Hamiltonian evolution unitary or the truncation of entanglement between qubits in the Matrix Product State simulator \cite{PhysRevX.10.041038,PRXQuantum.4.020304,PhysRevA.106.052430,Noh2020efficientclassical}. In the previous version of the pre-print the results provided for \qcels{} at $28$ qubits were based on data that was strongly damped and so was not well fit by our function of three complex exponentials with a dominant frequency of the ground state energy and instead a high frequency fit with a period of approximately the distance between data points was found, giving the appearance of a reasonable result. In figure \ref{fig:28qbs_faulty}, the decay of the oscillation can be observed in the data as well as the erroneous fit. In figure \ref{fig:28qbs_sort_of_fixed}, we forbid a fit with a period smaller than the spacing of the data, however, the decay of the oscillations continues to prevent a good fit to the data. In comparison, the adapt-QSCI algorithm continues to work well up to $28$ qubits, with its formulation as a classical solver acting in a subspace defined from the measurements on a quantum computer being well suited to working at moderate system sizes. It also works well at mitigating the impact of inexact quantum evolution from noise on a quantum device or error in the classical simulation of the quantum algorithm.

\section{Conclusion}
\label{sec:conc}

In this paper, we presented interdisciplinary work that merges software engineering and machine learning paradigms to enhance quantum algorithms' performance on NISQ hardware. This idea explored using Hybrid Quantum-Classical Systems as a promising way to utilise current NISQ hardware. While classical computers are essential to optimise noisy quantum hardware, the workload balance should still keep the quantum device at the core of the computation for such systems to stay relevant. The limit between optimising the quantum circuit outcome and bypassing the hardware limitations with classical means is difficult to define; it is however important to keep this balance in mind when developing hybrid systems.

We designed a new framework, implemented as a prototype tool, \tool, to predict quantum algorithm (near to) optimal hyperparameters. We evaluated \tool on two implementations, training and deploying relatively small-scale models to improve performance by suggesting better hyperparameters. Our results raised the possibility that the model's ability to predict optimal hyperparameters depends on the Hamiltonian characteristics rather than solely on a specific implementation (\adapt{} or \qcels).

%%%%%%%%%%%%%%%%%%%%%%%%%%%%%%%%%%%%%%%%%%%%%%%%%%%%%%%%%%%%%%%%%%
%\subsection{Discussion}
%\ab{Avner's idea.}
%\all{Workload Classical / Quantum computer}
%Some discussion:
%\paragraph{Computational Workload}
%Hybrid Quantum - Classical Systems are a promising way to utilise current NISQ hardware. While classical computers are essential to optimise noisy quantum hardware, the workload balance should still keep the quantum device at the core of the computation for such systems to stay relevant. The limit between optimising the quantum circuit outcome and bypassing the hardware limitations with classical means is difficult to define; it is however important to keep this balance in mind when developing hybrid systems.

%We work in two departments - say.
%- Tried ideas from CS that are not known to QP.
%- From Q side is a new way of opt things, and CS size is a new application case.
%Yet there is tension. Explain both sides high level.

%From QP side challenge:
%Restictions do not  make any sense.
%We think to do A, B, and C later.
%- ML training
%- Spliting the Phases. + parallel the process.
%- the rest of the ml things.

%From CS side: (we have no idea how things really works, or access to test it)
%- Too expensive.
%- No access to real QC. Maybe we get the results soon?

\vspace{1em}
\noindent\textbf{Code and Data Availability. }
The code, the training data, the models and the results are available as open-source at \cite{bensoussan_2024_artifact_ML2Q,bensoussan_2024_artifact_ML2QarXiv}.

\vspace{1em}
\noindent\textbf{Acknowledgments. } 
Authors are listed in alphabetical order.
We thank CloudLab \cite{cloudlab2019} for the platform and infrastructure support that enabled the experiment, resulting in the construction of two models for accelerating quantum algorithms. Sophie Fortz, Connor Lenihan and Avner Bensoussan are partially supported by the  EPSRC project on Verified Simulation for Large Quantum Systems (VSL-Q), grant reference EP/Y005244/1. Sophie Fortz and Avner Bensoussan are partially funded by the EPSRC project on Robust and Reliable Quantum Computing (RoaRQ), Investigation 009 - Model-based monitoring and calibration of quantum computations (ModeMCQ), grant reference EP/W032635/1. Sophie Fortz is partially funded by the QAssure project from Innovate UK. Also, King’s Quantum grants provided by King’s College London are gratefully acknowledged.
\bibliographystyle{abbrvnat}
\bibliography{main}

\end{document}